\documentclass[aps, twocolumn, letterpaper]{revtex4}

\usepackage{amsmath}
\usepackage{amssymb}
\usepackage{mathrsfs}
\usepackage{xspace}
\usepackage{graphicx}
\usepackage{braket}

\usepackage{ifpdf}
\ifpdf
\fi

\newcommand{\eq}[1]{(\ref{#1})}
\newcommand{\Eq}[1]{Eq.~(\ref{#1})}
\newcommand{\Eqs}[1]{Eqs.~(\ref{#1})}

\newcommand{\Sec}[1]{Sec.~\ref{#1}}
\newcommand{\Ref}[1]{Ref.~\cite{#1}}
\newcommand{\Refs}[1]{Refs.~\cite{#1}}
\newcommand{\App}[1]{Appendix~\ref{#1}}

\newcommand{\eg}{{e.g.,\/}\xspace}				
\newcommand{\ie}{{i.e.,\/}\xspace}				

\newcommand{\pd}{\partial}
\newcommand{\del}{\vec{\nabla}}
\newcommand{\cc}{\text{c.\,c.}}
\newcommand{\hc}{\text{h.\,c.}}

\newcommand{\mc}[1]{\mathcal{#1}}
\newcommand{\mcc}[1]{\mathfrak{#1}}
\newcommand{\msf}[1]{\mathsf{#1}}
\newcommand{\mcu}[1]{\mathscr{#1}}
\newcommand{\oper}[1]{\hat{\vec{#1}}}
\renewcommand{\vec}[1]{{\boldsymbol{\rm #1}}}
\newcommand{\favr}[1]{\langle #1 \rangle}
\newcommand{\msection}[1]{\textit{#1.}\ ---\ }

\usepackage{color} 

\begin{document}

\title{Lagrangian geometrical optics of nonadiabatic vector waves and spin particles}

\author{D.~E. Ruiz and I.~Y. Dodin}
\affiliation{Department of Astrophysical Sciences, Princeton University, Princeton, New Jersey 08544, USA}

\begin{abstract}
Linear vector waves, both quantum and classical, experience polarization-driven bending of ray trajectories and polarization dynamics that can be interpreted as the precession of the ``wave spin''. Both phenomena are governed by an effective gauge Hamiltonian, which vanishes in leading-order geometrical optics. This gauge Hamiltonian can be recognized as a generalization of the Stern-Gerlach Hamiltonian that is commonly known for spin-$1/2$ quantum particles. The corresponding reduced Lagrangians for continuous nondissipative waves and their geometrical-optics rays are derived from the fundamental wave Lagrangian. The resulting Euler-Lagrange equations can describe simultaneous interactions of $N$ resonant modes, where $N$ is arbitrary, and lead to equations for the wave spin, which happens to be a $(N^2-1)$-dimensional spin vector. As a special case, classical equations for a Dirac particle ($N = 2$) are deduced formally, without introducing additional postulates or interpretations, from the Dirac quantum Lagrangian with the Pauli term. The model reproduces the Bargmann-Michel-Telegdi equations with added Stern-Gerlach force.
\end{abstract}

\maketitle

\bibliographystyle{full}

\section{Introduction}

The geometrical-optics (GO) approximation is widely used to model waves in very diverse contexts, which range from quantum particle dynamics to electromagnetic (EM), acoustic, and even gravitational phenomena \cite{book:tracy, book:whitham, my:amc, ref:isaacson68}. As it is well known, GO is a theory asymptotic with respect to a small parameter, $\epsilon$, that is a ratio of the wave relevant characteristic period (temporal or spatial) to the inhomogeneity scale of the underlying medium. Absent resonances, just the lowest-order (``eikonal'' \cite{book:tracy}, or ``$\epsilon^0$'') approximation is applied most commonly, within which a wave exactly matches one of the local eigenmodes at any given location; then vector waves can be treated exactly like scalar waves \cite{my:amc}. However, this $\epsilon^0$ model misses important physics when $\epsilon$ is not vanishingly small or, alternatively, at large enough times. By including the first-order corrections in GO, one finds polarization-driven bending of ray trajectories \cite{ref:littlejohn91,Liberman:1992bz, Onoda:2004ij,ref:bliokh08,foot:stone, Liu:2015jj}, and one can also describe both adiabatic and diabatic mode conversion at resonances \cite{ref:kravtsov07, ref:bliokh07}. These $\epsilon^1$-effects were studied extensively in applications to quantum particles and EM waves \cite{foot:history}. However, the existing theories remain \textit{ad~hoc} (\eg are restricted to transverse waves in media with no spatial dispersion) and also cannot describe resonant coupling of more than two modes, if at all. Hence, they cannot treat many effects that are both relevant and important, particularly those in warm plasmas \cite{tex:myautozen}. The existing $\epsilon^1$-theories also have not quite unified relativistic quantum and EM effects beyond qualitative analogies and simple cases \cite{book:dragoman, ref:zapasskii99}. Those are long-standing problems \cite{foot:history}; however, the recent development of a universal axiomatic description of nondissipative linear waves \cite{my:wkin} makes them potentially solvable.

Here, we develop a general first-principle $\epsilon^1$-theory of resonant nondissipative vector waves. We start with the fundamental representation of the wave Lagrangian density (LD), \Eq{eq:gl}, decompose the field into local eigenmodes, and then simplify the resulting LD by neglecting terms of the second and higher orders in $\epsilon$. The reduced LD that is obtained describes some $N$ eigenmodes of the underlying medium, which are coupled through an effective ``gauge Hamiltonian'' of order $\epsilon$. We consider waves with frequencies in a narrow enough range, $O(\epsilon)$, around the central frequency; in this case, the gauge Hamiltonian can be expressed as a function of coordinates and the central wave vector. As such, it serves as a generalization of the Stern-Gerlach (SG) Hamiltonian that is commonly known for spin-$1/2$ quantum particles. We then show how to parameterize the LD in order to obtain dynamical equations for continuous GO waves and also for their rays. These equations describe both adiabatic and diabatic mode conversion of all $N$ resonant modes simultaneously. We also show that the dynamics of the $N$-dimensional complex polarization vector can be represented as the precession of a real $(N^2-1)$-dimensional fixed-length vector, which is interpreted as the wave spin. (For the special case when $N = 2$, a similar vector has been known as pseudospin, or Stokes vector \cite{ref:kravtsov07, ref:bliokh07}.) 

To illustrate our theory, we also apply it to derive a classical model of a Dirac particle, which by itself has been a long-standing and controversial problem; see \Refs{ref:gaioli98, arX:heinemann96, ref:ternov80, tex:rohrlich72, ref:plahte67} for reviews. We show that our formulation leads to the well known Bargmann-Michel-Telegdi (BMT) equations \cite{ref:bargmann59, foot:jackson} but with added SG terms, which are typically missed in straightforward semiclassical theories \cite{ref:rubinow63, ref:spohn00}. Our calculation is the first one that \textit{formally deduces} these corrected BMT equations from quantum theory without postulating any quantum-classical correspondence except the GO limit. (A more detailed comparison with earlier theories is given in \Sec{sec:comparison}.) This part of the presented research also can be considered as a follow-up to our \Ref{tex:myqlagr}, where the correspondence between quantum and classical LDs was studied in application to spinless and Pauli particles. 

Overall, the advantages of our new theory are as follows: (i)~The theory is derived in a variational form, so the resulting equations are manifestly conservative. (ii)~The theory assumes no specific wave equation; hence, the equivalence between spin effects in quantum and classical waves is automatically made quantitative. (iii)~The theory is naturally suited to serve as a stepping stone for studying ponderomotive effects on vector waves, in continuation of the recent \Ref{my:lens}, where studies of such forces on scalar waves were initiated. In particular, it renders possible a first-principle calculation of spin corrections to ponderomotive forces on electrons, which problem has been enjoying much attention recently \cite{ref:stefan11, arX:andreev14c, ref:brodin10, arX:wen14, ref:raicher13, ref:raicher14}. However, reporting this and other applications of the formalism presented here is left to future publications.
 
The paper is organized as follows. In \Sec{sec:notation}, we define the basic notation. In \Sec{sec:basic}, we present a general formalism describing vector waves. In \Sec{sec:rmodel}, we derive a general expression for the reduced LD of a wave, but its specific parameterizations are left to the follow-up sections. In \Sec{sec:scalar}, we discuss the leading-order approximation, in which the LD is parameterized as a scalar wave and leads to the standard equations of leading-order GO. In \Sec{sec:perturb}, we discuss a more precise model yet treat polarization effects as a perturbation. In \Sec{sec:self}, we present self-consistent fluid and point-particle models. In \Sec{sec:dirac}, we apply our theory to a Dirac particle and compare our model with related theories. In \Sec{sec:discussion}, we place our results in the wider context of general wave studies. In \Sec{sec:conclusion}, we summarize our main results.

\section{Notation}
\label{sec:notation}

The following notation is used throughout the paper. The symbol ``$\doteq$'' denotes definitions, ``$\cc$'' and ``\hc'' denote ``complex conjugate'' and ``Hermitian conjugate'', respectively; also, $\mathbb{I}_N$ denotes a unit $N \times N$ matrix, and hat ($\hat{\hphantom{k}}$) is reserved for differential operators. We use natural units, so the speed of light equals one ($c = 1$), and so is the Planck constant ($\hbar = 1$). The Minkowski metric is adopted with signature $(-, +, +, +)$, so, in particular, $\mathrm{d}^4x \equiv \mathrm{d}t\,\mathrm{d}^3x$. Generalizations to curved metrics are straightforward to apply \cite{my:wkin}. Greek indexes span from $0$ to $3$ and refer to spacetime coordinates, $x^\mu$, with $x^0$ corresponding to the time variable, $t$; in particular, $\pd_\mu \equiv \pd/\pd x^\mu$. Latin indexes span from $1$ to $3$ and denote the spatial variables, $x^i$ (except where specified otherwise); in particular, $\pd_i \equiv \pd/\pd x^i$. Summation over repeated spatial indexes is assumed. (However, the summation rule does not apply to mode indexes $q$ and $r$.) Also, for a given matrix $M$, we define its Hermitian and anti-Hermitian parts as
\begin{gather}\notag
M_H \doteq \frac{1}{2}\,(M + M^\dag), \quad M_A \doteq \frac{1}{2i}\,(M - M^\dag),
\end{gather}
so $M = M_H + i M_A$. Note that $iM_A$ is anti-Hermitian while $M_A$ itself is Hermitian. Also, in Euler-Lagrange equations (ELEs), the notation ``$\delta a: $'' denotes that the corresponding equation was obtained by extremizing the action integral with respect to $a$. Finally, the abbreviations used in the text are summarized as follows:\\[10pt]
\begin{tabular}{@{\quad} r@{\quad -- \quad} l @{\quad}}
ACT & action conservation theorem,\\
BMT & Bargmann-Michel-Telegdi,\\
ELE & Euler-Lagrange equation, \\
EM & electromagnetic, \\
GO & geometrical optics,\\
LD & Lagrangian density,\\
SG & Stern-Gerlach.\\
\end{tabular}\\

\section{Basic equations}
\label{sec:basic}

\subsection{Lagrangian density and Hamiltonian}

The dynamics of any nondissipative wave is governed by the least action principle, $\delta \Lambda = 0$, where $\Lambda$ is the action integral,
\begin{gather}\label{eq:actint}
\Lambda = \int \mcc{L}\,\mathrm{d}^4x.
\end{gather}
For a linear wave, the function $\mcc{L}$, termed LD, always can be written in the following representation \cite{my:wkin},
\begin{gather}\label{eq:gl}
\mcc{L} = \frac{i}{2}\,[\psi^\dag (\pd_t \psi) - (\pd_t \psi^\dag)\psi] - \psi^\dag \hat{H} \psi,
\end{gather}
which we call (the density of) the fundamental wave Lagrangian. Here $\psi$ is a complex vector field (``state function'') of some dimension $\bar{N}$, $\psi^\dag$ is its adjoint, and $\hat{H}$ is some Hermitian operator called Hamiltonian. In the Minkowski space assumed here, it also can be expressed as $\hat{H} = H(t, \oper{x}, \oper{k})$, where $H$ is some $\bar{N} \times \bar{N}$ matrix function, $\oper{x} = \vec{x}$ is the position operator, $\oper{k} = -i \del$ is the wave vector (momentum) operator, and the standard coordinate representation is assumed~\cite{my:wkin}.

We will consider $\hat{H}$ linear in~$\oper{k}$. In some systems, such as a Dirac particle (\Sec{sec:dirac}), Hamiltonians have this form originally; in others, $\hat{H}$ can be made linear in~$\oper{k}$ by extending the state function (\App{app:lin}) or by expanding the true Hamiltonian around some large central wave vector \cite{my:wkin}. In any case, the most general representation of such (Hermitian) $\hat{H}$ can be adopted in the form
\begin{gather}\label{eq:aux1}
\hat{H} = \mc{A}^0_H + (\vec{\mc{A}} \cdot \oper{k})_H
\end{gather}
where $\mc{A}^\mu = \mc{A}^\mu(t, \vec{x})$ are some $\bar{N} \times \bar{N}$ matrices, and $\vec{\mc{A}}$ is a column comprised of $\mc{A}^j$. [In principle, terms of higher orders in $\oper{k}$ could be retained too, as a perturbation, but only if they remain less or comparable to the small energies to be discussed below, such as $U$ given by \Eq{eq:Udef}.] Notice now that
\begin{align}
(\vec{\mc{A}} \cdot \oper{k})_H 
& = \frac{1}{2}\,(\vec{\mc{A}} \cdot \oper{k} + \oper{k} \cdot \vec{\mc{A}}^\dag) \notag \\
& = \frac{1}{2}\,[(\vec{\mc{A}}_H + i\vec{\mc{A}}_A) \cdot \oper{k} + \oper{k} \cdot (\vec{\mc{A}}_H - i\vec{\mc{A}}_A)] \notag\\
& = \frac{1}{2}\,(\vec{\mc{A}}_H \cdot \oper{k} + \oper{k} \cdot \vec{\mc{A}}_H) + \frac{i}{2}\,[\mc{A}_A^j, -i\pd_j] \notag \\
& = \frac{1}{2}\,(\vec{\mc{A}}_H \cdot \oper{k} + \oper{k} \cdot \vec{\mc{A}}_H) - \mc{C}.
\end{align}
Here $[\cdot\,, \cdot]$ is a commutator, and
\begin{gather}
\mc{C} \doteq \frac{1}{2}\,[\pd_j, \mc{A}_A^j] = \frac{1}{2}\,\del \cdot \vec{\mc{A}}_A
\end{gather}
is a Hermitian matrix, as $i\vec{\mc{A}}_A$ is anti-Hermitian. Then,
\begin{gather}\label{eq:LH}
\mcc{L} = \frac{i}{2}\,[\psi^\dag (\pd_t \psi) - (\pd_t\psi^\dag) \psi] - \psi^\dag \hat{H} \psi.
\end{gather}
Here we introduced a Hermitian operator
\begin{gather}\label{eq:H}
\hat{H} \doteq \frac{1}{2}\,[\vec{\alpha} \cdot (-i \del) + (-i \del) \cdot \vec{\alpha}] + \lambda,
\end{gather}
which serves as a new Hamiltonian, and also the following Hermitian matrices:
\begin{gather}
\alpha^j \doteq \mc{A}_H^j, \quad \lambda \doteq \mc{A}_H^0 - \mc{C}.
\end{gather}

In addition to \Eq{eq:LH}, the LD also has other equivalent representations. Below, we present some of them that we will need for our purposes.

\subsection{Super-Hamiltonian $\mc{H}$}

First, it is convenient to rewrite \Eq{eq:LH} as follows:
\begin{gather}
\mcc{L} = -\psi^\dag (- i\pd_t + \hat{H})\psi -\pd_t (i\psi^\dag \psi/2).
\end{gather}
Since the latter term yields zero contribution to the action integral \eq{eq:actint}, the LD can be equivalently expressed~as
\begin{gather}
\mcc{L} = - \psi^\dag\hat{\mc{H}}\psi, \quad \hat{\mc{H}} \doteq -i \pd_t + \hat{H}.
\end{gather}
The operator $\hat{\mc{H}}$ can be understood as the super-Hamiltonian and can be reexpressed as follows,
\begin{multline}
\hat{\mc{H}} 
= \frac{1}{2}\,[\alpha^\mu (-i\pd_\mu) + (- i\pd_\mu) \alpha^\mu] + \lambda \\
= \alpha^\mu (-i\pd_\mu) + \lambda - \frac{i}{2}\,(\del \cdot \vec{\alpha}),
\end{multline}
where $\alpha^0 \doteq \mathbb{I}_{\bar{N}}$. The corresponding ELEs are
\begin{gather}
\delta \psi^\dag : \quad \hat{\mc{H}} \psi = 0
\end{gather}
and the adjoint equation for $\psi^\dag$.

\subsection{Alternative form of $\boldsymbol{\mcc{L}}$}

Equivalently, \Eq{eq:LH} can be cast as follows:
\begin{align}
\mcc{L}
= \frac{i}{2}\,[&\psi^\dag (\pd_t\psi) - (\pd_t \psi^\dag)\psi] + \frac{i}{2}\,\psi^\dag \vec{\alpha} \cdot \del \psi  \notag \\
  & + \frac{i}{2}\,\psi^\dag \del \cdot (\vec{\alpha} \psi) - \psi^\dag \lambda \psi  \notag \\
= \frac{i}{2}\,[&\psi^\dag (\pd_t\psi) - (\pd_t\psi^\dag) \psi] + \frac{i}{2}\,[\psi^\dag \vec{\alpha} \cdot (\del \psi) \notag \\
 	& - (\del \psi^\dag) \cdot \vec{\alpha} \psi] - \psi^\dag \lambda \psi
 	+ \frac{i}{2}\, \del \cdot (\psi^\dag \vec{\alpha} \psi).
\end{align}
Like before, one can omit the divergence term; then,
\begin{multline}
\mcc{L}
= \frac{i}{2}\,[\psi^\dag (\alpha^\mu \pd_\mu) \psi - \cc]-\psi^\dag \lambda \psi \\
= - \text{Re}\,[\psi^\dag (-i \alpha^\mu \pd_\mu \psi +\lambda) \psi].\label{eq:LReH}
\end{multline}
This result also flows from $\mcc{L} = - \text{Re}\,(\psi^\dag \hat{\mc{H}} \psi)$, since ${\psi^\dag i(\del \cdot \vec{\alpha}) \psi}$ is imaginary due to $\del \cdot \vec{\alpha}$ being Hermitian.

\subsection{Problem outline}

Below, we will consider waves such that
\begin{gather}\label{eq:xi}
\psi = e^{i \theta} \xi,
\end{gather}
where $\theta$ is some rapid real phase (yet to be specified), and $\xi$ is a vector evolving slowly compared to $\theta$. This implies the following: (i)~$\alpha$ and $\lambda$ evolve slowly (if at all), and (ii)~among all the $\bar{N}$ dispersion branches, there are only some $N \le \bar{N}$ branches that are excited, which have local frequencies in some narrow enough range, $\Delta \omega/\omega = O(\epsilon) \ll 1$, around the central frequency
\begin{gather}\label{eq:omega}
\omega \doteq - \pd_t \theta.
\end{gather}
We will call these branches ``active'' and, for clarity, assign to them indexes $q = 1, \ldots, N$. [The remaining, ``passive'' branches will be assigned indexes $q = (N + 1), \ldots, \bar{N}$, correspondingly.] The goal of the calculation presented below is to derive an approximate LD for the active modes that would be accurate up to $O(\epsilon^1)$.

\section{Reduced model}
\label{sec:rmodel}

\subsection{Exact eigenmode representation}

In general, there exist $\bar{N}$ eigenfrequencies $\omega^q(k)$ corresponding to a given local wave vector
\begin{gather}\label{eq:k}
\vec{k} \doteq \del \theta.
\end{gather}
Those are found from the local dispersion relation,
\begin{gather}
\det [H(\vec{k}) - \mathbb{I}_{\bar{N}}\omega^q(\vec{k})] = 0,\\
H(\vec{k}) \doteq \vec{\alpha} \cdot \vec{k} + \lambda.
\end{gather}
(The dependence on $t$ and $\vec{x}$ is present too but will not be emphasized except when it is necessary.) Corresponding to $\omega^q(\vec{k})$ are some eigenvectors of $H(\vec{k})$, which we denote as $h_q$. Since $H(\vec{k})$ is Hermitian, $h_q$ can be chosen such that they form an orthonormal basis, in which any $\xi$ can be decomposed as
\begin{gather}
\xi = \sum_{q = 1}^{\bar{N}} h_q \bar{a}^q.
\end{gather}
The coefficients $\bar{a}^q$ are scalar complex functions of $x^\nu$. Typically, $\bar{a}_q^\dag = \left(\bar{a}^q\right)^*$, but we prefer to distinguish $\bar{a}_q^\dag$ and $\left(\bar{a}^q\right)^*$ to allow also for (somewhat exotic) negative-energy waves, which have $\bar{a}_q^\dag = -(\bar{a}^q)^*$ and should not be confused with negative-frequency waves \cite{foot:nege}. This also renders the notation more compact.

Now let us express the connection between $\xi$ and $\bar{a}^q$ in a matrix form, $\xi = \bar{\Xi} \bar{a}$. Here $\bar{a}$ is a column vector (of length $\bar{N}$) comprised of $\bar{a}^q$, and $\bar{\Xi}$ is a $\bar{N}\times \bar{N}$ fundamental matrix, which has vectors $h_q$ as its columns. Also let us adopt the notation $\mc{I} \doteq \psi^\dag \psi$ and $\bar{\mc{W}} \doteq \psi^\dag H(\vec{k}) \psi$ (these two functions are identified as the wave action density and the energy density, respectively \cite{my:wkin}),
\begin{gather}\notag
\mc{I} 
= \sum_{q, r = 1}^{\bar{N}} \bar{a}_q^\dag h^q h_r\bar{a}^r 
=\sum_{q, r = 1}^{\bar{N}} \bar{a}_q^{\dag} \delta_q^r \bar{a}^r 
= \sum_{q = 1}^{\bar{N}} \bar{a}_q^\dag \bar{a}^q
= \bar{a}^\dag \bar{a},
\end{gather}
\begin{multline}\notag
\bar{\mc{W}}
= \sum_{q, r = 1}^{\bar{N}} \bar{a}_q^\dag h^q H(\vec{k}) h_r \bar{a}^r
= \sum_{q, r = 1}^{\bar{N}} \bar{a}_q^\dag h^q\omega^r h_r \bar{a}^r \\
= \sum_{q, r = 1}^{\bar{N}} \bar{a}_q^\dag \omega^r \delta_q^r \bar{a}^r
= \sum_{q = 1}^{\bar{N}} \omega^q \bar{a}_q^\dag \bar{a}^q.
\end{multline}
Then, \Eq{eq:LReH} yields
\begin{align}
\mcc{L}
& = - \text{Re}\,[\psi^\dag \left(- i\pd_t - i \vec{\alpha} \cdot \del + \lambda \right) \psi]  \notag \\
& = - \mc{I}\,\pd_t \theta - \bar{\mc{W}} - \text{Re}\,[\bar{a}^\dag \bar{\Xi}^\dag \alpha^\mu (-i \pd_\mu) (\bar{\Xi} \bar{a})] \notag \\
& = - \mc{I}\,\pd_t \theta - \bar{\mc{W}} - \text{Im}\,[\bar{a}^\dag \bar{\Xi}^\dag \alpha^\mu \pd_\mu(\bar{\Xi} \bar{a})] \notag \\
& = - \mc{I}\,\pd_t \theta - \bar{\mc{W}} + \bar{\mc{K}},
\end{align}
where we introduced
\begin{multline}
\bar{\mc{K}} \doteq 
\frac{i}{2}\,[\bar{a}^\dag (\bar{\Xi}^\dag \alpha^\mu \bar{\Xi}) \pd_\mu \bar{a} - \cc] \\
+ \frac{i}{2}\,[\bar{a}^{\dag} \bar{\Xi}^\dag \alpha^\mu (\pd_\mu \bar{\Xi})\bar{a} - \cc].
\end{multline}

\subsection{Reduced eigenmode representation}

Since passive modes have nonzero amplitudes only due to the inhomogeneity of the medium, one has
\begin{gather}
\bar{a}^q = \left\{ 
 \begin{array}{ll}
  O(\epsilon^0),        & q = 1, \ldots, N \\
  O(\epsilon^1), & q = (N + 1), \ldots, \bar{N} \\
 \end{array}.
\right.
\end{gather}
Then, up to an error $O(\epsilon^2)$, one gets
\begin{gather}
\mc{I} \approx \sum_{q = 1}^N a_q^\dag a^q = a^\dag a, \\
\bar{\mc{W}} \approx \sum_{q = 1}^N \omega^q a_q^\dag a^q = a^\dag \mc{E} a,
\end{gather}
where $a^q \doteq \bar{a}^q$ for $q = 1, \ldots, N$, $a$ is a column vector comprised of all $a^q$, and $\mc{E}$ is a diagonal $N \times N$ matrix with eigenvalues $\omega^q$,
\begin{gather}
\mc{E} \doteq \text{diag}\, (\omega^1, \omega^2, \ldots, \omega^N).
\end{gather}

It is sufficient to keep only the active-mode contribution also in $\bar{\mc{K}}$, since both $\pd_\mu \bar{a}$ and $\pd_\mu \bar{\Xi}$ are already of order $\epsilon$ or less. Thus, within the accuracy $O(\epsilon^1)$, passive modes do not contribute to $\mcc{L}$ at all and can be omitted entirely. Instead of $\bar{\Xi}$, we hence can use its corresponding projection, $\Xi$, which is a $\bar{N} \times N$ (\ie nonsquare) matrix that has first $N$ vectors $h_q$ as its columns. This gives
\begin{gather}\label{eq:aux5}
\mcc{L} = - \mc{I}\,\pd_t\theta + \mc{K} - a^\dag (\mc{E} - U) a,
\end{gather}
where
\begin{gather}
\mc{K} \doteq \frac{i}{2}\,[a^\dag (\Xi^\dag \alpha^\mu \Xi) (\pd_\mu a) - (\pd_\mu a^\dag) (\Xi^\dag \alpha^\mu \Xi)a],\\
U \doteq \frac{i}{2}\,[\Xi^\dag \alpha^\mu (\pd_\mu \Xi) - (\pd_\mu \Xi^\dag) \alpha^\mu \Xi].\label{eq:Udef}
\end{gather}

\subsection{Group velocity and convective derivative}

To calculate $\mc{K}$ and $U$ with an error less than $O(\epsilon^1)$, it is enough to calculate $\Xi$ within the accuracy $O(\epsilon^0)$. Then, the difference in the active-mode frequencies can be neglected (i.e., $\omega^q \approx \omega$), and $\Xi$ approximately satisfies
\begin{gather}\label{eq:aux3}
0 = \mc{H}_0 \Xi \equiv (\vec{\alpha} \cdot \vec{k} + \lambda - \mathbb{I}_{\bar{N}} \omega) \Xi.
\end{gather}
Differentiating \Eq{eq:aux3} with respect to $\vec{k}$ gives an equation for the central group velocity, $v_0^i \doteq \pd \omega/\pd k_i$; namely,
\begin{gather}
\alpha^i \Xi + (\vec{\alpha} \cdot \vec{k} + \lambda -\mathbb{I}_{\bar{N}} \omega) \frac{\pd \Xi}{\pd k_i} - v_0^i \Xi = 0,
\end{gather}
which leads to
\begin{gather}\label{eq:aux4}
(\mathbb{I}_N v_0^i-\alpha^i) \Xi = \mc{H}_0 \frac{\pd \Xi}{\pd k_i}.
\end{gather}

Let us multiply \Eq{eq:aux4} by $\Xi^\dag$. Due to orthonormality of $h_q$, we can substitute ${\Xi^\dag \Xi = \mathbb{I}_N}$, and, due to \Eq{eq:aux3}, we also can substitute $\Xi^\dag \mc{H}_0 = 0$. That gives
\begin{gather}\label{eq:v0}
\mathbb{I}_N \vec{v}_0 = \Xi^\dag \vec{\alpha} \Xi
\end{gather}
(in which sense $\vec{\alpha}$ serves as the group velocity operator). Therefore,
\begin{gather}\notag
\Xi^\dag \alpha^\mu \Xi\, \pd_\mu = \Xi^\dag \Xi\, \pd_t + \Xi^\dag \vec{\alpha} \Xi \cdot \del = \mathbb{I}_N (\pd_t+ \vec{v}_0 \cdot \del) \equiv d_t
\end{gather}
can be understood as a convective derivative associated with velocity $\vec{v}_0$. [As will be shown below, $\vec{v}_0$ differs from the true average velocity, $\vec{V}$, by $O(\epsilon^1)$.] Hence we can simplify the expression for $\mc{K}$ as follows:
\begin{gather}
\mc{K} = \frac{i}{2}\,[a^\dag (d_t a)-(d_t a^\dag) a].
\end{gather}

\subsection{$\boldsymbol{U}$ as a function of $\boldsymbol{(t, \vec{x}, \vec{k})}$}
\label{sec:gauge}

Now let us search for a tractable expression for $U$. To do so, consider $\Xi$ as a function $\Xi(t, \vec{x}, \vec{k})$. (Below we assume these arguments by default.) Then,
\begin{gather}
\pd_\mu \Xi(t, \vec{x}, \vec{k}(t, \vec{x})) = \frac{\pd \Xi}{\pd x^\mu} + \frac{\pd \Xi}{\pd k_i} \, \pd_\mu k_i,
\end{gather}
where the former term on the right is understood as the derivative of $\Xi$ with respect to $x^\mu$ at fixed $\vec{k}$ (and $t$). Substituting this into \Eq{eq:Udef} yields
\begin{widetext}
\begin{multline}
U 
= \frac{i}{2} \left(\Xi^\dag \alpha^\mu \, \frac{\pd \Xi}{\pd x^\mu} - \frac{\pd \Xi^\dag}{\pd x^\mu} \, \alpha^\mu \Xi \right)
  + \frac{i}{2} \left[\Xi^\dag \alpha^\mu \, \frac{\pd \Xi}{\pd k_i} \left(\pd_\mu k_i\right)
  - \frac{\pd \Xi^\dag}{\pd k_i}\,(\pd_\mu k_i) \alpha^\mu \Xi^\dag\right] \\
= \frac{i}{2} \left(\Xi^\dag \alpha^\mu \, \frac{\pd \Xi}{\pd x^\mu} - \frac{\pd \Xi^\dag}{\pd x^\mu} \, \alpha^\mu \Xi \right)
  + \frac{i}{2} \left(\Xi^\dag \, \frac{\pd \Xi}{\pd k_i} - \frac{\pd \Xi^\dag}{\pd k_i} \, \Xi \right) \mathbb{I}_N \pd_t k_i
  + \frac{i}{2} \left(\Xi^\dag \alpha^j \, \frac{\pd \Xi}{\pd k_i} - \frac{\pd \Xi^\dag}{\pd k_i} \, \alpha^j \Xi \right) \pd_j k_i.
\end{multline}
Now recall that, by definition, $\vec{k}$ satisfies the so-called consistency relation,
\begin{gather}
\pd_t k_i 
= \pd_{i,t}^2 \theta 
= \pd_{t,i}^2 \theta 
= - \pd_i \omega(t, \vec{x}, \vec{k}(t, \vec{x}))
= - \frac{\pd \omega(t, \vec{x}, \vec{k})}{\pd x^i} - \frac{\pd \omega(t, \vec{x}, \vec{k})}{\pd k_j} \, \pd_i k_j.
\end{gather}
[Here and further, $\pd^2_{i,t} \equiv \pd^2/(\pd x^i \pd t)$, etc.] Using also that $\pd_i k_j = \pd_{j,i}^2 \theta = \pd_{i,j}^2 \theta = \pd_j k_i$, one gets
\begin{gather}
U 
= \frac{i}{2} \left(
\Xi^\dag \alpha^\mu \, \frac{\pd \Xi}{\pd x^\mu} 
- \frac{\pd \Xi^\dag}{\pd x^\mu} \, \alpha^\mu\Xi \right)
- \frac{i}{2} \left(\Xi^\dag \, \frac{\pd \Xi}{\pd k_i} 
- \frac{\pd \Xi^\dag}{\pd k_i} \, \Xi 
\right) 
\mathbb{I}_N \, \frac{\pd \omega(t, \vec{x}, \vec{k})}{\pd x^i} - \frac{i R}{2},
\end{gather}
\begin{align}
R
& \doteq \left(\Xi^\dag \frac{\pd \Xi}{\pd k_i} - \frac{\pd \Xi^\dag}{\pd k_i} \, \Xi \right)\mathbb{I}_N v_0^j \, \pd_{i,j}^2 \theta 
  - \left(\Xi^\dag \alpha^j \, \frac{\pd \Xi}{\pd k_i} - \frac{\pd \Xi^\dag}{\pd k_i} \, \alpha^j \Xi \right) \pd_{i,j}^2 \theta \notag \\
& = \left[\left(\Xi^\dag \, \frac{\pd \Xi}{\pd k_i} - \frac{\pd \Xi^\dag}{\pd k_i} \, \Xi \right) \mathbb{I}_N v_0^j
  - \left(\Xi^\dag \alpha^j \, \frac{\pd \Xi}{\pd k_i} - \frac{\pd \Xi^\dag}{\pd k_i} \, \alpha^j \Xi \right)\right] \pd_{i,j}^2 \theta \notag \\
& = \left[\left(\Xi^\dag \, \frac{\pd \Xi}{\pd k_i} \, \mathbb{I}_N v_0^j - \frac{\pd \Xi^\dag}{\pd k_j} \, \Xi  \mathbb{I}_N v_0^i\right)
  - \left(\Xi^\dag \alpha^j \, \frac{\pd \Xi}{\pd k_i} - \frac{\pd \Xi^\dag}{\pd k_j} \, \alpha^i \Xi \right)\right] \pd_{i,j}^2 \theta \notag \\
& = \left[\Xi^\dag (\mathbb{I}_N v_0^j-\alpha^j)\, \frac{\pd \Xi}{\pd k_i}
  - \frac{\pd \Xi^\dag}{\pd k_j} \, (\mathbb{I}_N v_0^i - \alpha^i) \Xi \right] \pd_{i,j}^2 \theta.
\end{align}
But, from \Eq{eq:aux4}, we know that
\begin{gather} \label{eq:aux12}
\Xi^\dag (\mathbb{I}_N v_0^j-\alpha^j) = \frac{\pd \Xi^\dag}{\pd k_j} \, \mc{H}_0,
\qquad
(\mathbb{I}_N v_0^i - \alpha^i) \Xi = \mc{H}_0 \, \frac{\pd \Xi}{\pd k_i}.
\end{gather}
Therefore, $R = 0$, so the resulting $U$ can be rewritten as follows:
\begin{align}
U
= & \frac{i}{2} \left(\Xi^\dag \alpha^\mu \, \frac{\pd \Xi}{\pd x^\mu}
  - \frac{\pd \Xi^\dag}{\pd x^\mu} \, \alpha^\mu \Xi \right)
  - \frac{i}{2} \left(\Xi^\dag \, \frac{\pd \Xi}{\pd k_i} - \frac{\pd \Xi^\dag}{\pd k_i} \, \Xi \right) \frac{\pd \omega(t, \vec{x}, \vec{k})}{\pd x^i} \notag \\
= & \frac{i}{2} \left(\Xi^\dag \, \frac{\pd \Xi}{\pd t} - \frac{\pd \Xi^\dag}{\pd t} \, \Xi \right)
  + \frac{i}{2} \left(\Xi^\dag \alpha^i \, \frac{\pd \Xi}{\pd x^i} - \frac{\pd \Xi^\dag}{\pd x^i} \, \alpha^i \Xi \right)
  - \frac{i}{2} \left(\Xi^\dag \, \frac{\pd \Xi}{\pd k_i} - \frac{\pd \Xi^\dag}{\pd k_i} \, \Xi \right) \frac{\pd \omega(t, \vec{x}, \vec{k})}{\pd x^i} \notag \\
= & \frac{i}{2} \left(\Xi^\dag \, \frac{\pd \Xi}{\pd t} - \frac{\pd \Xi^\dag}{\pd t} \, \Xi \right)
  + \frac{i}{2} \left[\left(\Xi^\dag v_0^i - \frac{\pd \Xi^\dag}{\pd k_i} \, \mc{H}_0 \right) \frac{\pd \Xi}{\pd x^i}
  - \frac{\pd \Xi^\dag}{\pd x^i} \left(\Xi v_0^i - \mc{H}_0 \, \frac{\pd \Xi}{\pd k_i}\right)\right]
  - \frac{i}{2} \left(\Xi^\dag \, \frac{\pd\Xi}{\pd k_i} - \frac{\pd \Xi^\dag}{\pd k_i} \, \Xi \right) \frac{\pd \omega(t, \vec{x}, \vec{k})}{\pd x^i} \notag \\
= & \frac{i}{2} \left(\Xi^\dag \, \frac{\pd \Xi}{\pd t} - \hc \right)
  + \frac{i}{2} \left(\Xi^\dag v_0^i \, \frac{\pd \Xi}{\pd x^i} - \hc \right)
  - \frac{i}{2} \left(\Xi^\dag \, \frac{\pd \Xi}{\pd k_i} - \hc \right) \frac{\pd \omega(t, \vec{x}, \vec{k})}{\pd x^i}
  - \frac{i}{2} \left(\frac{\pd \Xi^\dag}{\pd k_i} \, \mc{H}_0 \, \frac{\pd \Xi}{\pd x^i} - \hc \right) \notag \\
= & \frac{i}{2} \left[\Xi^\dag \left(\pd_t + \vec{v}_0 \cdot \del - \del \omega \cdot \pd_{\vec{k}}\right) \Xi - \hc \right]
  + \frac{i}{2} \left(\frac{\pd \Xi^\dag}{\pd x^i} \, \mc{H}_0 \, \frac{\pd \Xi}{\pd k_i} - \hc \right),\label{eq:U2}
\end{align}
\end{widetext}
where we substituted corollaries of \Eqs{eq:aux12},
\begin{gather}\notag
\Xi^\dag \alpha^i = \mathbb{I}_N v_0^i - \frac{\pd \Xi^\dag}{\pd k_j} \, \mc{H}_0, 
\quad 
\alpha^i \Xi = \mathbb{I}_N v_0^i - \mc{H}_0 \, \frac{\pd \Xi}{\pd k_i}.
\end{gather}

Equation \eq{eq:U2} can be rewritten more compactly as follows. Notice that $\vec{v}_0$ and $- \del \omega$ are components of the phase flow velocity in the ray phase space, as determined by the zeroth-order (in $\epsilon$) ray equations,
\begin{gather}
\dot{\vec{x}}(t) = \pd_{\vec{k}} \omega (t, \vec{x}, \vec{k}),
\quad
\dot{\vec{k}}(t)=-\pd_{\vec{x}}\omega (t,\vec{x},\vec{k}).
\end{gather}
Thus, it is convenient to introduce the convective derivative in the ray phase space, or the Liouville operator,
\begin{multline}
\mathbb{D}_t \Xi \doteq \frac{d}{dt}\, \Xi (t, \vec{x}(t), \vec{k}(t))
\\ = \left[\pd_t + \dot{\vec{x}}(t) \cdot \pd_{\vec{x}} + \dot{\vec{k}}(t) \cdot \pd_{\vec{k}} \right] \Xi (t, \vec{x}, \vec{k}).
\end{multline}
Then, \Eq{eq:U2} becomes
\begin{multline}
U(t, \vec{x}, \vec{k}) = \frac{i}{2} \left[\Xi^\dag (\mathbb{D}_t \Xi) - (\mathbb{D}_t \Xi)^\dag \Xi \right]
  \\ + \frac{i}{2} \left(
    \frac{\pd \Xi^\dag}{\pd \vec{x}} \cdot \mc{H}_0 \cdot \frac{\pd \Xi}{\pd \vec{k}} 
    - \frac{\pd \Xi^\dag}{\pd \vec{k}} \cdot \mc{H}_0 \cdot \frac{\pd \Xi}{\pd \vec{x}}
    \right). \label{eq:U3}
\end{multline}

The importance of \Eq{eq:U3} is that it guarantees $U$ to be a function of $(t, \vec{x}, \vec{k})$ (and not of the gradient of $\vec{k}$, the dependence on which on has been eliminated when we proved that $R = 0$). Due to this property, $-U$ can be identified as an effective ``gauge Hamiltonian'' \cite{foot:gauge} with a well defined classical limit. In application to specific systems such as atoms and guided EM waves with simple dispersion, related potentials were also studied, for instance, in \Refs{arX:goldman13, ref:fang13b, ref:dalibard11}. In application to EM waves propagating in weakly inhomogeneous isotropic dielectric media, $U$ serves as the Hamiltonian of the interaction between the photon spin (polarization) and the photon orbital motion \cite{Liberman:1992bz, Onoda:2004ij, ref:bliokh07, ref:bliokh08}. Also, for EM waves in weakly anisotropic media, $U$ determines the interaction between the photon spin and the pseudo-magnetic field that is effectively caused by the medium's anisotropy \cite{ref:bliokh07, Liu:2015jj}. For waves and particles propagating in rapidly oscillating backgrounds \cite{my:lens}, $U$ is also recognized as minus the ponderomotive energy. Finally, for quantum particles, $U$ serves as (or is related to) the SG Hamiltonian. This will become clear from Secs.~\ref{sec:self} and \ref{sec:dirac}, after we discuss the ELEs that flow from the reduced Lagrangian derived here.

\subsection{Lagrangian density: summary}
\label{sec:sum}

Let us split the total new Hamiltonian, $\mc{E} - U$, into the average energy,
\begin{gather}\label{eq:H0def}
H_0(t, \vec{x}, \vec{k}) \doteq N^{-1} \text{Tr}(\mc{E} - U),
\end{gather}
and the following traceless matrix,
\begin{gather}
\Omega(t, \vec{x}, \vec{k}) \doteq \mc{E} - U - \mathbb{I}_N H_0.
\end{gather}
(Below, we omit $\mathbb{I}_N$ for brevity.) Then, \Eq{eq:aux5} can be rewritten as follows,
\begin{gather}\label{eq:L0}
\mcc{L} = \frac{i}{2}\,[a^\dag (d_t a) - (d_t a^\dag) a] - (\pd_t \theta + H_0)\, a^\dag a - a^\dag \Omega a.
\end{gather}
Since $\vec{k} \equiv \del \theta$, both $H_0$ and $\Omega$ are functions of $(t, \vec{x}, \del \theta)$, and so is $\vec{v}_0$; then,
\begin{gather}
d_t = \pd_t + \vec{v}_0 (t, \vec{x}, \del \theta) \cdot \del.
\end{gather}

Notice also that
\begin{align}
\frac{i}{2}\,[a^\dag (d_t a) - (d_t a^\dag) a]
= & \,\frac{i}{2}\,[a^\dag (\pd_t a) - (\pd_t a^\dag) a] \notag \\
	& + \frac{i}{2}\,[a^\dag \vec{v}_0 \cdot (\del a) - (\del a^\dag) \cdot \vec{v}_0 a] \notag \\
= & \,i a^\dag d_t a + \frac{i}{2}\,(\del \cdot \vec{v}_0) a^\dag a \notag \\
	& - \frac{i}{2}\, \pd_\mu (a^\dag v_0^\mu a), \notag
\end{align}
where $v_0^0 \doteq 1$. Since the divergence term can be dropped, this leads to the following equivalent form of the LD:
\begin{gather}\label{eq:L1}
\mcc{L} = a^\dag \left[i d_t + \frac{i}{2}\,(\del \cdot \vec{v}_0) - (\pd_t \theta + H_0) - \Omega \right] a.
\end{gather}

Equations \eq{eq:H0def}-\eq{eq:L1}, in combination with \Eqs{eq:Udef} and \eq{eq:U3} for $U$ and with \Eq{eq:v0} for $\vec{v}_0$, are the main results of this section. Below, we discuss how to apply these results in various cases.

\section{Scalar-wave limit}
\label{sec:scalar}

To the lowest, zeroth order in $\epsilon$, the LD can be approximated simply with
\begin{gather}
\mcc{L} = - (\pd_t\theta + H_0) a^\dag a = - (\pd_t\theta + H_0)\mc{I},
\end{gather}
where $H_0$ also can be replaced with $\mc{E}$. One may recognize this as the standard LD for a scalar wave in Hayes's form \cite{ref:hayes73}, which corresponds to the $\epsilon^0$-theory. Such LD is parameterized by just two independent functions, the rapid phase $\theta$ and the total action density $\mc{I}$. The corresponding ELEs are the action conservation theorem (ACT),
\begin{gather}\label{eq:act0}
\delta \theta : \quad \pd_t \mc{I} + \del \cdot (\mc{I}\vec{v}_0) = 0,
\end{gather}
and a Hamilton-Jacobi equation,
\begin{gather}
\delta \mc{I} : \quad \pd_t \theta + H_0(t, \vec{x}, \del \theta) = 0,\label{eq:sHJ}
\end{gather}
which can be understood as the local dispersion relation,
\begin{gather}
\omega = H_0(t, \vec{x}, \vec{k})
\end{gather}
[cf. \Eqs{eq:omega} and \eq{eq:k}]. The resulting dynamics do not need to be discussed here, since it is exhaustively covered in literature. For an overview, see, \eg \Refs{my:amc, book:tracy}.

\section{Prescribed $\boldsymbol{\theta}$}
\label{sec:perturb}

\subsection{Basic equations}

Now let us include terms in $\mcc{L}$ of the first order in $\epsilon$. Postponing a comprehensive discussion until \Sec{sec:self}, here we outline polarization effects in a simplified manner, namely, as a perturbation to the scalar dynamics described in \Sec{sec:scalar}. Within this approach, we can choose $\theta$ (which has been unspecified so far) as a prescribed function satisfying \Eq{eq:sHJ}. Then \Eq{eq:L1} becomes
\begin{gather}
\mcc{L} = a^\dag \left[id_t + \frac{i}{2}\,(\del \cdot \vec{v}_0) - \Omega \right] a,
\end{gather}
where the vectors $a$ and $a^\dag$ can be adopted as independent variables. (The latter would have not been possible if $\theta$ were itself to be found; see \Sec{sec:self}.) This leads to the following ELE,
\begin{gather}
\delta a^\dag : \quad i d_t a = \Omega a - \frac{i}{2}\,(\del \cdot \vec{v}_0)a,
\end{gather}
and the adjoint ELE for $a^\dag$. As a corollary, the following equation for $\mc{I}$ is yielded,
\begin{gather}
d_t \mc{I} = a^\dag d_t a + \cc = - (\del \cdot \vec{v}_0)\mc{I},
\end{gather}
which is equivalent to \Eq{eq:act0}. Hence, one can simplify the motion equation by introducing normalized variables,
\begin{gather}
b \doteq \mc{I}^{-1/2}a, \quad b^\dag b = 1.
\end{gather}
The new variables satisfy
\begin{gather}\label{eq:beq}
id_t b
= \mc{I}^{-1/2} id_t a - \frac{i d_t\mc{I}}{2\mc{I}}\, b = \Omega b.
\end{gather}
This equation shows that, in the frame traveling with velocity $\vec{v}_0$, the vector $b$ rotates with matrix frequency $\Omega$, and $\mc{W} \doteq a^\dag \Omega a$ serves as the mode-coupling energy. 

\subsection{Wave spin}
\label{sec:spin}

Let us also describe the rotation of $b$ as follows. As a traceless Hermitian matrix, $\Omega$ can be represented as a linear combination of $N^2-1$ generators $T_u$ of $\text{SU}(N)$, which are traceless Hermitian matrices, with some real coefficients $-W^u$ \cite{tex:anisovich}:
\begin{gather}
\Omega = - \sum_{u = 1}^{N^2 - 1} T_u W^u \equiv - \vec{T} \cdot \vec{W}.
\end{gather}
It is instructive to introduce the $(N^2-1)$-dimensional vector $\vec{S} \doteq b^\dag \vec{T} b$, so that $\mc{W} = \mc{I} \favr{\Omega}$, where $\favr{\Omega} \doteq - \vec{S} \cdot \vec{W}$. The components of $\vec{S}$ satisfy the following equation:
\begin{align}
d_t S_w 
& = b^\dag T_w (d_t b) + (d_t b^\dag) T_w b \notag \\
& = i b^\dag \Omega T_w b - i b^\dag T_w \Omega b \notag \\
& = i b^\dag (\Omega T_w - T_w \Omega)b \notag \\
& = i b^\dag [\Omega, T_w] b \notag \\
& = i \,b^\dag [T_w, T_u] b W^u \notag \\
& = - f_{wuv} (b^\dag T^v b) W^u, \notag \\
& = f_{wvu} S^v W^u, \label{eq:aux7}
\end{align}
where $f_{wuv}$ are structure constants, which are antisymmetric in all indexes \cite{tex:anisovich}. 

For example, consider the case when only two waves are resonant. Then, $N^2 - 1 = 3$, $T^v$ are the three Pauli matrices divided by two (so $|\vec{S}|^2 = 1/2$), and $f_{wuv}$ is the Levi-Civita symbol, so $f_{wvu} S^v W^u = (\vec{S} \times \vec{W})_w$. For a Dirac electron, which is a special case (\Sec{sec:dirac}), such $\vec{S}$ is recognized as the spin vector undergoing the well known precession equation, $d_t\vec{S} = \vec{S} \times \vec{W}$. (One may also recognize this as an equation for the Stokes vector that was derived earlier to characterize the polarization of transverse EM waves in certain media \cite{ref:bliokh08, ref:kravtsov07, ref:bliokh07}.) Hence, it is convenient to extend this quantum terminology also to $N$ waves. We will call the corresponding $(N^2 - 1)$-dimensional vector $\vec{S}$ a generalized spin vector and express $f_{wvu} S^v W^u$ symbolically as $(\vec{S} * \vec{W})_w$, where $*$ can be viewed as a generalized vector product. Then, \Eq{eq:aux7} is rewritten compactly in the following vector form,
\begin{gather}\label{eq:spineq}
d_t \vec{S} = \vec{S} * \vec{W},
\end{gather}
and is understood as a generalized precession equation.

\subsection{Applicability of the prescribed-$\boldsymbol{\theta}$ model}
\label{sec:applic}

In some cases, the prescribed-$\theta$ model can be entirely sufficient. For example, suppose both $H_0$ and $\Omega$ are constant. Suppose also that $\theta$ changes only along a single axis, $x$, and is independent of transverse coordinates, $\rho$. (An example might be the case of a charged quantum particle traveling parallel to a constant magnetic field.) Then $\mcc{L}$ can be integrated over $\rho$, so a Lagrangian linear density is yielded that describes the one-dimensional motion along $x$. Due to $\Omega$ being constant and small, $b$ oscillates at the same small rate everywhere and thus remains slow compared to $\theta$. Then the prescribed-$\theta$ model is valid indefinitely.

However, if $\Omega$ is inhomogeneous, phases of $b$ at different locations grow at different rates. Then, even though $\Omega$ is small, the assumption of slow $\xi$ is eventually violated, and hence the prescribed-$\theta$ model cannot be trusted. (See also \Sec{sec:comparison}.) Equations that remain accurate on larger time scales, therefore, can be derived only within a model where the rapid phase $\theta$ is calculated self-consistently. Such model is discussed below.

\section{Self-consistent $\boldsymbol{\theta}$}
\label{sec:self}

\subsection{Fluid model}
\label{sec:fluid}

Let us adopt $(\theta, \mc{I})$ as independent variables, like in \Sec{sec:scalar}. The remaining variable, $b$, can be parameterized with the $(N - 1)$ spherical angles $\zeta^r$ on the $N$-dimensional sphere and some $(N - 1)$ relative phases $\vartheta^q$ of individual $b^q$; however, the remaining, $N$th phase is not independent. To account for that, this time we define $\theta$ such that, for example, $a^N$ be real. This leads to a new set of normalized variables, $\eta$. (Like $b$, the vector $\eta$ is defined via $a = \mc{I}^{1/2}\eta$. However, we distinguish $b$ and $\eta$ because the underlying definitions of $a$ is different due to the fact that the corresponding $\theta$s are defined differently.) The components of $\eta$ can be parameterized as
\begin{gather}
\eta^q = \Phi^q (\zeta^1, \ldots, \zeta^{N - 1}) \times \left\{
\begin{array}{ll}
 e^{- i \vartheta^q}, & q < N \\
 1, & q = N \\
\end{array}\right.
.
\end{gather}
Here $\Phi^q$ are the well known real functions parameterizing the location of a point on a unit sphere,
\begin{align}
\Phi^1 & = \cos(\zeta^1), \\
\Phi^2 & = \sin(\zeta^1) \cos(\zeta^2), \\
\Phi^3 & = \sin(\zeta^1) \sin(\zeta^2) \cos(\zeta^3), \\
& \kern 5pt \vdots \notag \\
\Phi^{N - 1} & = \sin(\zeta^1) \ldots \sin(\zeta^{N-2}) \cos(\zeta^{N - 1}), \\
\Phi^N & = \sin(\zeta^1) \ldots \sin(\zeta^{N-2}) \sin(\zeta^{N - 1}),
\end{align}
so $\sum^N_{q = 1} (\Phi^q)^2 \equiv 1$. Notice that the LD in \Eq{eq:L0} is then expressed as
\begin{gather}
\mcc{L} = - \mc{I} (\pd_t \theta + H_0 + \favr{\Omega}) + \frac{i\mc{I}}{2}\,[\eta^\dag (d_t \eta) - (d_t \eta^\dag) \eta],
\end{gather}
or, more explicitly,
\begin{multline}
\mcc{L} = 
- \mc{I} \left[\pd_t\theta + H_0(t, \vec{x}, \del \theta) 
+ \favr{\Omega}(t, \vec{x}, \del \theta, \vartheta, \zeta)\right] \\
+ \mc{I} \sum_{q = 1}^{N - 1} \left[\Phi^q(\zeta)\right]^2 (\pd_t + \vec{v}_0 \cdot \del) \vartheta^q,
\end{multline}
where the derivatives of $\zeta^q$ canceled out. (Note that we use $\vartheta$ to denote the whole set of $N - 1$ variables $\vartheta^r$, and similarly for $\zeta$.) This leads to the following $2N$ ELEs.

\msection{Action conservation theorem} The first ELE is the ACT,
\begin{gather}\label{eq:actV}
\delta \theta : \quad \pd_t \mc{I} + \del \cdot (\mc{I}\vec{V}) = 0,
\end{gather}
which is a continuity equation for $\mc{I}$. The corresponding flow velocity is $\vec{V} = \vec{v} + \vec{u}$, where
\begin{gather}
\vec{v} \doteq \pd_{\vec{k}} H_0+ \favr{\pd_{\vec{k}}\Omega},\\
\vec{u} \doteq - \sum_{q = 1}^{N - 1} (\Phi^q)^2\, \frac{\pd \omega}{\pd \vec{k} \pd \vec{k}} \cdot \del \vartheta^q.
\end{gather}
Notably, one can also recast the latter formula as $\vec{u} = \vec{M}^{-1} \cdot \vec{\kappa}$, where $\vec{M}$ is understood as the mass tensor of a wave quantum, and $\vec{\kappa}$ is the wave vector of $\eta$; namely,
\begin{gather}
\vec{M}^{-1} \doteq \frac{\pd^2 \omega}{\pd \vec{k} \pd \vec{k}},\\
\vec{\kappa} \doteq -\frac{i}{2}\,[\eta^\dag (\del \eta)-(\del \eta)^\dag \eta] = -i \eta^\dag \del \eta.
\end{gather}

\msection{Hamilton-Jacobi equation} The second ELE, 
\begin{gather}
\delta \mc{I} :  \quad \pd_t\theta + H_0 + \favr{\Omega} -\sum_{q = 1}^{N - 1} \left(\Phi^q\right)^2d_t \vartheta^q =0,
\end{gather}
can be considered as a generalization of the Hamilton-Jacobi equation to vector waves.

\msection{Equation for $\vartheta$} Another set of $(N - 1)$ ELEs is
\begin{gather}
\delta  \zeta^r : \quad
\frac{\pd \favr{\Omega}}{\pd \zeta^r} - \sum_{q = 1}^{N - 1} \frac{\pd (\Phi^q)^2}{\pd \zeta^r} \, d_t \vartheta^q = 0.
\end{gather}

\msection{Equation for $\zeta$} Another set of $(N - 1)$ ELEs is
\begin{gather}\label{eq:chi}
\delta \vartheta^r : \quad \pd_t [\mc{I} (\Phi^r)^2] + \del \cdot [\mc{I} (\Phi^r)^2 \vec{v}_0] + \mc{I}\,\frac{\pd \favr{\Omega}}{\pd \vartheta^r} = 0.
\end{gather}
Each of these equations describes the evolution of the action of an individual ($r$th) mode, $\mc{I} \left(\Phi^r\right)^2$. By using \Eq{eq:actV}, one can rewrite \Eq{eq:chi} also as follows:
\begin{multline}
\mc{I} (\pd_t + \vec{V} \cdot \del) (\Phi^r)^2 + \del \cdot [\mc{I} (\Phi^r)^2 (\vec{v}_0-\vec{V})] \\
+ \mc{I}\, \frac{\pd \favr{\Omega}}{\pd \vartheta^r} = 0.
\end{multline}
Note that, in the case of a localized wave packet, averaging over the packet area eliminates the divergence term and predicts the advection of $(\Phi^r)^2$ at velocity $\vec{V}$.

Combined together, the $2N$ equations derived in this section can be viewed as a generalization of the classical spin-fluid equations that we earlier derived \cite{tex:myqlagr} for a Pauli particle ($N = 2$). The generalization consists of the fact that the new equations apply to general waves (\eg classical EM waves), as opposed to specific waves of quantum matter. In particular, the spin is a $(N^2 - 1)$-dimensional vector now, in contrast to the three-dimensional spin of a Pauli particle.

\subsection{Point-particle model and ray equations}
\label{sec:rays}

To the extent that a wave packet is well localized such that it is meaningful to describe its dynamics as the dynamics of the packet's geometrical center, the continuous-wave description developed above can be replaced with a simpler, point-particle model. In this case, one can approximate the action density with a delta function,
\begin{gather}
\mc{I}(t, \vec{x}) = \delta(\vec{x} - \vec{X}(t)),
\end{gather}
so the LD can be replaced with just a point-particle Lagrangian, $L \doteq \int \mcc{L}\,\mathrm{d}^3x$. Following the same approach as in \Ref{tex:myqlagr}, one obtains
\begin{multline}
L = \vec{P} \cdot \dot{\vec{X}} - H_0(t, \vec{X}, \vec{P}) - \favr{\Omega}(t, \vec{X}, \vec{P}, \vartheta, \zeta) \\
+ \sum_{q = 1}^{N - 1} \left[\Phi^q(\zeta)\right]^2 \dot{\vartheta}^q,
\end{multline}
where $\vec{P}(t) \doteq \del \theta(t, \vec{X}(t))$; the angles $\vartheta$ and $\zeta$ are also evaluated at $\vec{x} = \vec{X}(t)$. The corresponding ELEs are
\begin{align}
\delta \vec{P} : & \quad \dot{\vec{X}} = \pd_{\vec{P}} [H_0 + \favr{\Omega}], \label{eq:r1}\\
\delta \vec{X} : & \quad \dot{\vec{P}} = - \pd_{\vec{X}} [H_0 + \favr{\Omega}], \label{eq:r2}\\
\delta \zeta^r : & \quad \sum_{q = 1}^{N - 1} \frac{\pd (\Phi^q)^2}{\pd \zeta^r}\,\dot{\vartheta}^q - \frac{\pd \favr{\Omega}}{\pd \zeta^r} = 0, \label{eq:r3}\\
\delta \vartheta^r  : & \quad \sum_{q = 1}^{N - 1} \frac{\pd (\Phi^r)^2}{\pd \zeta^q}\,\dot{\zeta}^q + \frac{\pd \favr{\Omega}}{\pd \vartheta^r} = 0.\label{eq:r4}
\end{align}
[Notably, $\favr{\Omega}$ can be viewed as the Hamiltonian for $(\vartheta, \zeta)$, but the equations for $(\vartheta, \zeta)$ have a non-canonical form. In contrast, the equations for $\vec{X}$ and $\vec{P}$ are canonical.] These equations can also be considered as ray equations for the fluid equations derived in \Sec{sec:fluid}.

As before, an equation for the spin vector flows from the ELEs as a corollary. At least for a Pauli particle, this is easily shown by a straightforward calculation; see also \Ref{tex:myqlagr}. However, instead of rederiving the spin equation from scratch, let us propose the following argument.

\subsection{Complex representation}
\label{sec:crays}

Instead of $(\vartheta, \zeta)$, one can also use $(\eta^\dag, \eta)$ as independent variables, if one requires explicitly that $\eta$ be constrained by $\eta^\dag \eta = 1$ and $\eta_N^\dag = \eta^N$. The constraints can be implemented by introducing two Lagrange multipliers, $\mu$ and $i\nu$, so that the resulting Lagrangian becomes
\begin{multline}
L 
= \vec{P} \cdot \dot{\vec{X}} - H_0(t, \vec{X}, \vec{P}) + \frac{i}{2}\,(\eta^\dag \dot{\eta} - \dot{\eta}^\dag \eta) 
\\ - \eta^\dag \Omega(t, \vec{X}, \vec{P}) \eta + \mu(\eta^\dag \eta - 1) + i\nu(\eta_N^\dag - \eta^N). \label{eq:Lcons}
\end{multline}
The corresponding ELEs are as follows:
\begin{align}
\delta \eta_{q < N}^\dag & : \quad i\dot{\eta}^q = (\Omega \eta)^q + \mu \eta^q, \label{eq:e1}\\
\delta \eta_{q < N}      & : \quad - i\dot{\eta}_q^\dag = (\eta^\dag \Omega)_q + \eta_q^\dag \mu, \label{eq:e2}\\
\delta \eta_N^\dag       & : \quad i\dot{\eta}^N = (\Omega \eta)^N + \mu \eta^N - i\nu, \label{eq:e3}\\
\delta \eta_N            & : \quad -i\dot{\eta}_N^\dag = (\eta^\dag \Omega)_N + \eta_N^\dag \mu + i\nu, \label{eq:e4}\\
\delta \mu            & : \quad \eta^\dag \eta - 1 = 0, \label{eq:e5}\\
\delta \nu            & : \quad \eta_N^\dag - \eta^N = 0, \label{eq:e6} \\
\delta \vec{P}        & : \quad \dot{\vec{X}} = \pd_{\vec{P}} (H_0 + \eta^\dag \Omega \eta), \label{eq:e7} \\
\delta \vec{X}        & : \quad \dot{\vec{P}} = - \pd_{\vec{X}} (H_0 + \eta^\dag \Omega \eta). \label{eq:e8}
\end{align}
By comparing \Eqs{eq:e1} and \eq{eq:e2}, it is seen that $\mu$ must be real; then, \Eqs{eq:e4}-\eq{eq:e6} show that $\nu$ is real too \cite{foot:real}. Now let us introduce $\mc{U} \doteq \Omega + \mu \mathbb{I}_N$ to write
\begin{gather}
\dot{\eta}^q = - i(\mc{U} \eta)^q - \nu \delta_{q, N},
\end{gather}
where $\delta_{q, N}$ is the Kronecker symbol. This leads to
\begin{gather}
\frac{d}{dt} \, (\eta^\dag \eta) = - 2 \nu \eta^N,
\end{gather}
since $\mc{U}$ is Hermitian. But, from \Eq{eq:e5}, we know that $d(\eta^\dag \eta)/dt = 0$. Since $\eta^N$ cannot remain zero identically, this leaves us with $\nu = 0$, so the entire vector $\eta$ satisfies
\begin{gather}\label{eq:etaeq}
i \dot{\eta} = \mc{U} \eta.
\end{gather}

Notice, however, that these ELEs do not form a closed system, as there is no independent equation for $\mu$. The issue can be evaded by rewriting the ELEs in terms of
\begin{gather}\label{eq:zdef}
z(t) \doteq \eta(t) \exp\left(i \int^t \mu(t')\,\mathrm{d}t'\right).
\end{gather}
According to \Eq{eq:etaeq}, $z$ satisfies
\begin{gather}\label{eq:z}
i \dot{z} = \Omega z.
\end{gather}
We then obtain, like in \Sec{sec:spin}, that
\begin{gather}\label{eq:SB}
\dot{\vec{S}} = \vec{S} * \vec{W}, \quad \vec{S} \doteq z^\dag \vec{T} z.
\end{gather}
Equations \eq{eq:e7} and \eq{eq:e8} can be expressed through $\vec{S}$ too,
\begin{gather}
\dot{\vec{X}} = \pd_{\vec{P}} H_0 - \vec{S} \cdot \pd_{\vec{P}} \vec{W}, \label{eq:X2}\\
\dot{\vec{P}} = - \pd_{\vec{X}} H_0 + \vec{S} \cdot \pd_{\vec{X}} \vec{W}, \label{eq:P2}
\end{gather}
where $H_0$ and $\vec{W}$ are functions of $(t, \vec{X}, \vec{P})$. 

In contrast to the ELEs, \Eqs{eq:SB}-\eq{eq:P2} \textit{do} form a closed system. It is also seen now that the effect of polarization on the wave ray, or point-particle, dynamics is akin to that of the SG Hamiltonian on a Pauli particle \cite{tex:myqlagr} [except, in general, the vector $\vec{S}$ is a $(N^2 - 1)$-dimensional]. Also, it is to be noticed that this generalized SG Hamiltonian can contribute to the expression for the particle velocity, $\dot{\vec{X}}$, so the latter is not necessarily equal to the average group velocity, $\pd_{\vec{P}} H_0$.

Notice, finally, that the resulting equations for variables $(\vec{X}, \vec{P}, z^\dag, z)$ also can be assigned a Lagrangian,
\begin{multline}
L 
= \vec{P} \cdot \dot{\vec{X}} - H_0(t, \vec{X}, \vec{P}) 
\\ + \frac{i}{2}\,(z^\dag \dot{z} - \dot{z}^\dag z) - z^\dag \Omega(t, \vec{X}, \vec{P}) z,
\label{eq:Lz}
\end{multline}
assuming that the initial conditions are restricted to $z^\dag z = 1$. Specifically, the corresponding ELEs are
\begin{align}
\delta \vec{P} : & \quad \dot{\vec{X}} = \pd_{\vec{P}} H_0 - \vec{S} \cdot \pd_{\vec{P}} \vec{W}, \\
\delta \vec{X} : & \quad \dot{\vec{P}} = - \pd_{\vec{X}} H_0 + \vec{S} \cdot \pd_{\vec{X}} \vec{W}, \\ 
\delta z^\dag  : & \quad \dot{z} = -i\Omega z , \\
\delta z       : & \quad \dot{z}^\dag = i(\Omega z)^\dag
\end{align}
and yield the spin equation \eq{eq:SB} as a corollary. But keep in mind that \textit{these} equations [as opposed to \Eqs{eq:r1}-\eq{eq:r4}] do not allow finding the full vector $\eta$, because $\mu$ in the variable transformation \eq{eq:zdef} remains unknown.

\section{Example: Dirac particle}
\label{sec:dirac}

In this section, we apply the above formalism to a Dirac particle considered as an example. Our goal is to obtain a first-principle Lagrangian $\epsilon^1$-theory that describes Dirac particles as point particles with spin. In \Sec{sec:comparison}, we will also discuss how our theory relates to other existing theories of classical and semiclassical Dirac particles.

\subsection{Basic equations}

First, let us introduce the commonly known Dirac LD,
\begin{gather}
\mcc{L} = - \psi^\dag [\alpha^\mu (-i\pd_\mu - q A_\mu) + \beta m]\psi.
\end{gather}
Here $q$ and $m$ are the particle charge and mass (the assumed sign convention is that $q < 0$ for electrons), $A_\mu$ is the four-vector potential, $\alpha$ and $\beta$ are the corresponding Dirac matrices,
\begin{gather}\notag
\alpha^0 =
\begin{pmatrix}
\mathbb{I}_2 & 0 \\
0 & \mathbb{I}_2 
\end{pmatrix} 
, \quad
\vec{\alpha} =
\begin{pmatrix}
0 & \vec{\sigma} \\
\vec{\sigma} & 0 
\end{pmatrix} 
, \quad
\beta =
\begin{pmatrix}
\mathbb{I}_2 & 0 \\
0 & -\mathbb{I}_2
\end{pmatrix},
\end{gather}
$\mathbb{I}_2$ is a $2\times 2$ unit matrix, and $\vec{\sigma}$ are the Pauli matrices. (The Dirac gamma matrices and the Dirac adjoint do not need to be introduced for our purposes.) This $\mcc{L}$ can be cast in the form \eq{eq:LH}. This is done by adopting
\begin{gather}
\lambda = - q \vec{\alpha} \cdot \vec{A} + \beta m,
\end{gather}
while the Dirac alpha matrices serve precisely as the alpha matrices of our theory (but now they are constant). 

The total number of modes is $\bar{N} = 4$, and the frequencies $\omega^q$ (corresponding to the limit $\epsilon = 0$) are
\begin{align}
\omega^1 & = \omega^2 = \omega_+ \doteq + \varepsilon - q A_0, \\
\omega^3 & = \omega^4 = \omega_- \doteq - \varepsilon - q A_0.
\end{align}
Here $\varepsilon$ is the kinetic energy, introduced as
\begin{gather}
\varepsilon (p) \doteq \sqrt{m^2 + p^2},
\end{gather}
$\vec{p}$ is the kinetic momentum, introduced as
\begin{gather}
\vec{p} \doteq \vec{P} - q \vec{A},
\end{gather}
and the standard notation $\vec{P} \equiv \vec{k}$ is introduced to denote the canonical momentum. (In our notation, $\vec{P}$ and $\vec{k}$ do not need to be distinguished because $\hbar = 1$.)

\subsection{Effective potential}

For clarity, let us consider the case where there are only two active modes ($N = 2$), specifically, those corresponding to spin-up and spin-down particle; hence, no antiparticles are considered. Then, as it is well known~\cite{foot:stone},
\begin{gather}\label{eq:Xi0}
\Xi = \sqrt{\frac{m + \varepsilon}{2\varepsilon}}\left(
\begin{array}{c}
 \mathbb{I}_2 \\
 \frac{\vec{\sigma} \cdot \vec{p}}{m + \varepsilon} \\
\end{array}
\right).
\end{gather}
The effective potential $U$ can be calculated straightforwardly using \Eq{eq:Udef}. To calculate $\pd_\mu \Xi$ entering that equation, let us consider $\Xi$ as a function
\begin{gather}
\Xi(t, \vec{x}) = \Xi (\varepsilon(t, \vec{x}), \vec{p}(t, \vec{x})).
\end{gather}
Then, $U = - (\mc{P}_t + \mc{P}_x + \mc{Q}_t + \mc{Q}_x)$, where (\App{app:auxdirac})
\begin{align}
\mc{P}_t & \doteq \text{Im} \left(\Xi^\dag \, \frac{\pd \Xi}{\pd p_j}\, \pd_t p_j\right) 
                =  \frac{\vec{v} \times \pd_t \vec{p}}{2(m + \varepsilon)}  \cdot \vec{\sigma}, \\
\mc{P}_x & \doteq \text{Im} \left(\Xi^\dag \alpha^i \, \frac{\pd \Xi}{\pd p_j}\, \pd_i p_j\right) 
                = -  \frac{q\vec{B}}{2\varepsilon} \cdot \vec{\sigma} ,\\
\mc{Q}_t & \doteq \text{Im} \left(\Xi^\dag \, \frac{\pd \Xi}{\pd \varepsilon} \, \pd_t \varepsilon \right) = 0,\\
\mc{Q}_x & \doteq \text{Im} \left(\Xi^\dag \alpha^i \, \frac{\pd \Xi}{\pd \varepsilon} \, \pd_i\varepsilon \right) 
                = \frac{\vec{v} \times \del \varepsilon}{2(m + \varepsilon)} \cdot \vec{\sigma}.
\end{align}
Here, $\vec{E}$ and $\vec{B}$ are the electric and magnetic fields, and $\vec{v} \doteq \vec{p}/\varepsilon$ is the particle velocity (unperturbed by the spin coupling). Since
\begin{align}
\del \varepsilon
& \approx \del \left(- \pd_t \theta + q A_0\right) \notag\\
& = - \pd_t \del \theta + q\del A_0 \notag\\
& = - \pd_t (\del \theta - q \vec{A}) - q \pd_t \vec{A} + q \del A_0 \notag\\
& = - \pd_t \vec{p} + q\left(-\pd_t \vec{A} + \del A_0\right) \notag\\
& = - \pd_t \vec{p} + q \vec{E},
\end{align}
one can rewrite $\mc{Q}_x$ also as follows,
\begin{gather}
\mc{Q}_x
= - \frac{\vec{v} \times \pd_t \vec{p}}{2(m + \varepsilon)} \cdot \vec{\sigma}
  + \frac{q \vec{v} \times \vec{E}}{2(m + \varepsilon)} \cdot \vec{\sigma}.
\end{gather}
Then, the terms proportional to $\pd_t \vec{p}$ cancel out (as expected from \Sec{sec:gauge}), and one arrives at
\begin{gather}
U = - \vec{\sigma} \cdot \left[- \frac{q \vec{B}}{2\varepsilon} + \frac{q\vec{v} \times \vec{E}}{2(m + \varepsilon)}\right]
  \equiv \frac{1}{2}\,\vec{\sigma} \cdot \vec{W},
\end{gather}
where we introduced
\begin{gather} \label{eq:W}
\vec{W} \doteq \frac{q\vec{B}}{\varepsilon} - \frac{q\vec{v} \times \vec{E}}{m + \varepsilon}
= \frac{q}{m \gamma}\left(\vec{B} - \frac{\gamma}{\gamma + 1}\,\vec{v} \times \vec{E} \right),
\end{gather}
and $\gamma \doteq \varepsilon/m$ is the particle Lorentz factor (unperturbed by the spin coupling). Since $U$ is traceless, we obtain
\begin{gather}\label{eq:OmH0}
\Omega = - U = - \frac{1}{2}\,\vec{\sigma} \cdot \vec{W}, \quad H_0 = \omega_+.
\end{gather}
The corresponding fluid and point-particle equations are hence derived as described in Secs.~\ref{sec:perturb} and \ref{sec:self}. In particular, the point-particle spin vector satisfies
\begin{gather}
\dot{\vec{S}} = \vec{S} \times \vec{W},
\end{gather}
so $\vec{W}$, which is given by \Eq{eq:W}, serves as the spin precession frequency. This agrees with the Thomas precession equation \cite{ref:thomas27, foot:rose}. Also note that, \Eq{eq:L0} with \Eq{eq:OmH0} taken in the nonrelativistic limit yields
\begin{multline}\notag
\mcc{L} = \frac{i}{2} \left[a^\dag (d_t a) - (d_t a^\dag) a \right] - (\pd_t \theta) a^\dag a\\
- a^\dag \left[\frac{1}{2m}\,(\del \theta - q\vec{A})^2 - q A_0 - \frac{q}{2m}\,(\vec{\sigma} \cdot \vec{B})\right] a.
\end{multline}
Substituting here $a = \mc{I}^{1/2}\eta$ leads to precisely the LD of a classical Pauli particle \cite{tex:myqlagr}. Thus, it also subsumes the classical limit of the Takabayasi equations \cite{ref:takabayasi55}, as discussed in \Ref{tex:myqlagr}.

\subsection{Anomalous magnetic moment}

Now let us also consider the correction due to an anomalous magnetic moment. This correction is described by adding to $\mcc{L}$ the Pauli term \cite{ref:salamin93, ref:gaioli98},
\begin{gather}\label{eq:dL}
\pounds = - \psi^\dag \left[\frac{q}{4m} \left(\frac{g}{2} - 1\right) \beta \sigma^{\mu\nu} F_{\mu\nu}\right] \psi.
\end{gather}
Here $g$ is the $g$-factor, $\sigma^{\mu\nu} \doteq (i/2)[\gamma^\mu, \gamma^\nu]$ is the relativistic spin operator, $\gamma^\mu \doteq \beta \alpha^\mu$ are the Dirac gamma matrices, and $F_{\mu\nu}$ is the EM field tensor. Then,
\begin{gather}
\sigma^{\mu\nu}F_{\mu\nu} =
2
\begin{pmatrix}
- \vec{\sigma} \cdot \vec{B} & i  \vec{\sigma} \cdot \vec{E}  \\
i \vec{\sigma} \cdot \vec{E} & -  \vec{\sigma} \cdot \vec{B}  \\
\end{pmatrix},
\end{gather}
which leads to (\App{app:diracan})
\begin{multline}\notag
\psi^\dag \beta \sigma_{\mu\nu} F^{\mu\nu} \psi \approx 
	- 2 a^\dag
	\bigg[
	\vec{\sigma} \cdot \vec{B}  - (\vec{v} \times \vec{E}) \cdot \vec{\sigma} \\
	\left. - \frac{\gamma}{\gamma + 1} \, (\vec{\sigma} \cdot \vec{v}) (\vec{B} \cdot \vec{v})
	\right] a.
\end{multline}
Substituting this into \Eq{eq:dL}, we get
\begin{gather}\notag
\pounds = \frac{q}{2m}\left(\frac{g}{2} - 1\right)\, a^\dag \vec{\sigma} a \cdot 
		\bigg[ \vec{B} - \vec{v}\times \vec{E}
		- \frac{\gamma}{\gamma + 1}\, (\vec{B} \cdot \vec{v}) \vec{v}\bigg].
\end{gather}
Then, we can express the contribution of the gauge Hamiltonian to the LD as
\begin{gather}
a^\dag U a + \pounds = a^\dag U_g a,
\end{gather}
where we introduced $U_g \doteq \vec{\sigma} \cdot \vec{W}_g/2$ and 
\begin{multline}\notag
\vec{W}_g \doteq 
		\frac{q}{m} \left[
 			\left(\frac{g}{2} - 1 + \frac{1}{\gamma} \right) \vec{B}
 			- \left(\frac{g}{2} - \frac{\gamma}{\gamma + 1}\right) \vec{v} \times \vec{E} \right. \\
 			\left. - \left(\frac{g}{2} - 1\right) \frac{\gamma}{\gamma + 1}\, (\vec{B} \cdot \vec{v}) \vec{v}
		\right].
\end{multline}

It is seen that the effect of an anomalous magnetic moment is described by replacing $U$ with $U_g$ or, in other words, by replacing $\vec{W}$ with $\vec{W}_g$; hence, the corrected spin precession frequency is $\vec{W}_g$. This leads to the well known BMT equation for $\vec{S}$ \cite{ref:bargmann59, foot:jackson}, which is thereby seen to be subsumed by our theory as a special case. Yet, in contrast to the BMT theory, our formulation also captures the SG force [for example, see \Eqs{eq:X2} and \eq{eq:P2}] and thus remains manifestly conservative also in the presence of field gradients.

\subsection{Comparison with other models}
\label{sec:comparison}

The \textit{sample application} of our theory to a Dirac particle discussed above can be viewed as a complement to the many spin-particle models yielded by an almost century-long research. The relevant literature is too extensive to be surveyed here, so we refer the reader to already existing reviews, \eg \Refs{ref:gaioli98, arX:heinemann96, ref:ternov80, tex:rohrlich72, ref:plahte67}. That said, let us briefly outline how our study fits the general context.

\msection{1} On the score of being manifestly Lagrangian, our theory is reminiscent of that by Barut \textit{et~al} \cite{ref:barut84, ref:barut85, ref:barut90, ref:barut93}. However, in contrast to ours, the latter (i)~is constructed axiomatically, rather than by deduction, (ii)~does not capture the SG force (at least, manifestly), (iii)~is solely a point-particle theory, rather than also a fluid theory, (iv)~and relies on the concept of proper time, which is introduced \textit{ad~hoc} and in a debatable manner \cite{foot:debat}. Also importantly, Barut \textit{et~al}'s theory (as well as many other classical-spin theories) is formulated in terms of bispinors, as opposed to spinors. We believe that this is an unnecessary complication. A general bispinor describes a superposition of a particle and an antiparticle. But the slow-envelope approximation generally is not possible for such an object; then the very concept of a point particle remains undefined. Moreover, a particle and an antiparticle respond very differently to external fields, so they do not travel as a whole. Thus, a point-particle model for them may not be physically meaningful, even when it can be constructed formally \cite{foot:kinfl}. Our theory avoids this issue by being explicitly restricted to pure (\ie particle \textit{or} antiparticle) states. Then the internal state is described by two rather than four complex numbers, which also simplifies the analysis.

\msection{2} In application to a Dirac particle specifically, the separation of pure states that we used can be interpreted as the Foldy-Wouthuysen transformation \cite{ref:foldy50} expanded asymptotically in $\epsilon$. In this sense, our calculation is also related to that in, \eg \Refs{ref:derbenev73, arX:heinemann96, ref:chen14}, but with the following difference. In contrast to other papers, and as we already pointed out in a different context, our theory is constructed straightforwardly by deduction from the quantum LD. We do not postulate additional symmetries, Poisson brackets, or any other correspondence between quantum and classical dynamics except the GO limit. The state function $\psi$ does not need to be related to probability in our formulation \cite{foot:prob}, and thus expectation values do not need to be introduced. (The same applies to Weyl symbols, which fact also simplifies the derivation conceptually.) In other words, \textit{our results are formal and do not require an interpretation}.

\msection{3} As a GO theory, our formulation is related to the many existing semiclassical theories of Dirac particles (for instance, see \Refs{ref:rubinow63, ref:spohn00}), but it is different from those theories. First, typical semiclassical calculations assume $\theta$ to satisfy the zeroth-order Hamilton-Jacobi equation \eq{eq:sHJ} and thus do not extend beyond what we call the prescribed-$\theta$ model. As explained in \Sec{sec:applic}, this model is generally inadequate at large times, and that is the reason why standard semiclassical expansions fail to capture the SG force. A related discussion can be found, for example, in \Refs{tex:rohrlich72, ref:rubinow63}. Second, our theory is Lagrangian and, as such, leads to manifestly conservative equations. The Lagrangian formulation is also what allows us to introduce a classical theory unambiguously (see also \Ref{tex:myqlagr}) and keep the equations manifestly conservative. Moreover, the general theory presented in this paper can be readily reformulated to treat also rapidly oscillating quasiperiodic fields, in which case $U$ becomes a ponderomotive potential. For that, one needs only to replace the assumed \Eq{eq:Xi0} for the zeroth-order fundamental matrix, $\Xi$, with a more appropriate expression such as a Volkov solution \cite{ref:bergou80, ref:salamin93}. We leave reporting such calculations to future publications. 

\msection{4} Finally, and most importantly, our theory is not restricted to Dirac particles. It is applicable, in fact, to general waves, both quantum and classical. Below we discuss this in more detail.

\section{Discussion}
\label{sec:discussion}

A related discussion where polarization effects are explored for general waves can be found in \Ref{ref:littlejohn91}. An expression for the energy akin to our $\Omega$ [with $U$ given by \Eq{eq:U3}] also appeared there. However, there are important differences between our work and \Ref{ref:littlejohn91}, which are as follows. First, the physical meaning of the mentioned term is not identified in \Ref{ref:littlejohn91}; the term is said to have ``no name''. In contrast, we identify $\Omega$ unambiguously as the (generalized) SG Hamiltonian, when we study Dirac and Pauli particles as a special case. Second, and most importantly, \Ref{ref:littlejohn91} is restricted to adiabatic dynamics. This means that the action is assumed to be conserved in each mode individually, so the analysis in \Ref{ref:littlejohn91} is inapplicable to systems with degenerate spectrum, including Dirac particles. For the same reason, the many existing studies \cite{foot:stone} on the Berry phase effect in the context of a Dirac particle (and beyond) are not directly related to the subject of our paper.

Nonadiabatic dynamics, which includes mode conversion, was studied, \eg in \Refs{ref:littlejohn92, ref:littlejohn93}, but in the restricted context of an asymptotic scattering problem and only for two-component waves \cite{foot:2x2}. In contrast, our theory introduces mode coupling directly in ray equations (Secs.~\ref{sec:rays} and \ref{sec:crays}) and describes simultaneous resonant interaction of arbitrarily many modes. (The latter and the manifestly Lagrangian formulation also distinguish our work from the already mentioned studies \cite{ref:kravtsov07, ref:bliokh07} of transverse EM waves in relatively simple media.) This is important, for example, for an adequate description of autoresonant wave-wave interactions \cite{tex:myautozen}. Other applications of the general theory could be in describing nonadiabatic polarization dynamics in nonstationary inhomogeneous plasmas or other media for which the existing theories are insufficient.

Although wave Hamiltonians $\hat{H}$ may not be known explicitly in such applications, our theory still can be applied as follows. Suppose a set of $\bar{N}$ linear equations of the first order in $\pd_\mu$. (For example, in plasma physics, those can be linearized hydrodynamic equations combined with Maxwell's equations.) Then, to the zeroth order in $\epsilon$, the equation for all field variables combined as a vector $\Psi$ is readily cast in the form
\begin{gather}\label{eq:Psi}
(\mcu{A} \cdot \vec{k} + \mcu{B} - \omega \mathbb{I}_{\bar{N}}) \Psi = 0.
\end{gather}
Now suppose that we are dealing specifically with nondissipative waves. By definition of such a wave \cite{my:wkin}, there is a variable transformation $\Psi \mapsto \psi$ that casts \Eq{eq:Psi} in the form \eq{eq:aux3} \cite{my:wkin}, \ie with Hermitian $\vec{\alpha}$ and $\lambda$. Once those matrices are found, one can find also $\Xi$, $\vec{v}_0$, $U$, and $\mcc{L}$ (see also Sec.~7.5 in \Ref{my:wkin}). Furthermore, the underlying formulation can be extended to a \textit{multifluid} description of vector waves; then it becomes applicable, for example, to improving ray tracing in plasmas. Discussions of these and other applications of the general theory presented here are left to future publications.

\section{Conclusions}
\label{sec:conclusion}

In summary, we developed a general first-principle \mbox{$\epsilon^1$-theory} of resonant nondissipative vector waves. We start with deriving the reduced LD that describes coupling of arbitrary $N$ eigenmodes in weakly nonstationary and inhomogeneous medium. The coupling term can be understood as an effective gauge Hamiltonian of the order of the GO parameter, $\epsilon$. As such, this gauge Hamiltonian serves as a generalization of the SG Hamiltonian that is commonly known for spin-$1/2$ quantum particles. We show how to parameterize the LD in order to obtain dynamical equations for continuous GO waves and also for their rays. These equations describe both adiabatic and diabatic mode conversion of all $N$ resonant modes simultaneously. We also show that the dynamics of the $N$-dimensional complex polarization vector can be represented as the precession of a real $(N^2-1)$-dimensional fixed-length vector, which is interpreted as the wave spin. As an example, we apply our theory to derive a classical model of a Dirac particle. We show that our formulation leads to the well known BMT equations but with added SG energy terms. Our calculation is the first one that \textit{formally deduces} these corrected BMT equations from quantum theory without postulating any quantum-classical correspondence except the GO limit. 

Overall, the advantages of the proposed theory are (i) its variational form, leading to manifestly conservative equations, and (ii) the fact that the theory assumes no specific wave equation and thus treats classical and quantum waves on the same footing. Also, the new theory is naturally suited to serve as a stepping stone for studying ponderomotive effects on vector waves and particles. Reporting this and other applications of the general formalism presented here is left to future publications.

The authors thank J.~W. Burby and N.~J. Fisch for valuable discussions. The work was supported by the NNSA SSAA Program through DOE Research Grant No. DE274-FG52-08NA28553, by the U.S. DOE through Contract No. DE-AC02-09CH11466, and by the U.S. DOD NDSEG Fellowship through Contract No. FA9550-11-C-0028.


\appendix

\section{Linearization of Hamiltonians with respect to $\boldsymbol{\hat{k}}$}
\label{app:lin}

\subsection{Hamiltonian representation}

Consider a Schr\"{o}dinger equation,
\begin{gather}\label{eq:PsiSE}
i\pd_t \Psi =\hat{\msf{H}}\Psi,
\end{gather}
where the Hamiltonian $\hat{\msf{H}}$ is a matrix polynomial of $\hat{k} \doteq -i \pd_x$ of some order $n > 1$. (We assume
one-dimensional coordinate space for simplicity, but generalizations to multidimensional spaces are straightforward to apply.) Since the Hamiltonian
is Hermitian, it can be expressed as
\begin{gather}\label{eq:msfH}
\hat{\msf{H}} = (\mc{A}_0 + \mc{A}_1 \hat{k} + \ldots + \mc{A}_n \hat{k}^n)_H,
\end{gather}
where $\mc{A}_m = \mc{A}_{m,H} + i \mc{A}_{m,A}$ are matrix functions of $x$. 

Let us focus on the term that is of the highest order in $\hat{k}$. It can be written as
\begin{gather}
(\mc{A}_n \hat{k}^n)_H = \hat{\mc{Q}}_n + \hat{\mc{R}}_n,
\end{gather}
where we introduced the following Hermitian operators,
\begin{gather}
\hat{\mc{Q}}_n \doteq \frac{1}{2}\,(\mc{A}_{n,H} \hat{k}^n + \hat{k}^n \mc{A}_{n,H}),
\\
\hat{\mc{R}}_n \doteq \frac{i}{2}\,[\mc{A}_{n,A}, \hat{k}^n],
\end{gather}
and $[\cdot\,, \cdot]$ is a commutator. Notice that $\hat{\mc{R}}_n = o(\hat{k}^n)$, where $o(\hat{k}^n)$ denotes an operator such that, when applied to $\Psi$, it contains derivatives of $\Psi$ only of orders less than $n$. Then, $\hat{\msf{H}} = \hat{\mc{Q}}_n + o(\hat{k}^n)$, where we used the fact that the remaining terms in \Eq{eq:msfH} are also $o(\hat{k}^n)$.

As a Hermitian matrix, $\mc{A}_{n,H}$ can be expressed as $\mc{A}_{n,H} = U^\dag D U$, where $U$ is unitary, and $D$ is diagonal with real eigenvalues. Then, $\hat{\msf{H}} = U^\dag \hat{K} U + o(\hat{k}^n)$, where
\begin{gather}
\hat{K}\doteq \frac{1}{2} \, (D \hat{k}^n + \hat{k}^n D)
\end{gather}
is diagonal. Let us represent the eigenvalues of $D$ as $D_q = \Sigma_q (d_q)^n$, where $\Sigma_q = \text{sgn}\, D_q$, and $d_q > 0$. Then,
\begin{gather}\notag
\hat{K} 
= \frac{1}{2}\,\text{diag}_q \{D_q \hat{k}^n + \hat{k}^n D_q \}
= \text{diag}_q \{\hat{G}_q^n \Sigma_q \} + o(\hat{k}^n)
\end{gather}
[$\text{diag}_q\{X_q\}$ denotes a diagonal matrix with eigenvalues $X_q$], where we introduced the scalar Hermitian operators
\begin{gather}
\hat{G}_q \doteq \frac{1}{2} \, (d_q\hat{k} + \hat{k} d_q).
\end{gather}
Introducing (commuting) matrices $\hat{G} \doteq \text{diag}_q\{\hat{G}_q\}$ and $\Sigma \doteq \text{diag}_q\{\Sigma_q\}$, one can further rewrite $\hat{K}$ compactly~as
\begin{gather}
\hat{K} = \hat{G}^n \Sigma + o(\hat{k}^n).
\end{gather}

Now consider a new variable, $\psi \doteq U \Psi$. Using \Eq{eq:PsiSE} and the fact that $U U^\dag$ is a unit matrix, we obtain a Schr\"{o}dinger
equation for $\psi$,
\begin{gather}\label{eq:efp}
i \pd_t\psi = \hat{H} \psi,
\end{gather}
where $\hat{H}$ is a Hermitian operator given by
\begin{gather}
\hat{H} = \tilde{H} + \hat{G}^n \Sigma, \quad \tilde{H} = o(\hat{k}^n).
\end{gather}

\subsection{Phase space extension}

Let us introduce an auxiliary function
\begin{gather}\label{eq:chidef}
\chi \doteq \mu^{-1} \hat{G}\Sigma \psi,
\end{gather}
where $\mu$ is a real constant parameter that is yet to be specified. In combination with \Eq{eq:chidef}, \Eq{eq:efp} can be represented as the following \textit{pair} of equations,
\begin{align}
i \pd_t\psi & = \tilde{H}\psi + \mu \hat{G}^{n-1}\chi, \label{eq:aux301}\\
          0 & = \mu^2 \hat{Y} (\mu^{-1} \hat{G} \Sigma \psi - \chi),\label{eq:aux302}
\end{align}
where the operator $\mu^2 \hat{Y}$ must be invertible but otherwise can be chosen arbitrarily. Equations \eq{eq:aux301} and \eq{eq:aux302} can be considered as a special case of the vector equation
\begin{gather}i \pd_t\left(
\begin{array}{c}
 \psi  \\
 \chi  \\
\end{array}
\right)=\left(
\begin{array}{cc}
 \tilde{H} & \mu  \hat{G}^{n-1} \\
 \mu \hat{Y} \hat{G} \Sigma  & -\mu^2 Y \\
\end{array}
\right)\left(
\begin{array}{c}
 \psi  \\
 \chi  \\
\end{array}
\right) \label{eq:chieq}
\end{gather}
in the limit $\mu \to \infty$. Finally, choosing $\hat{Y} = \hat{G}^{n-2} \Sigma$ turns \Eq{eq:chieq} into a Schr\"{o}dinger equation
with a manifestly Hermitian Hamiltonian operator of order $(n-1)$ in $\hat{k}$:
\begin{gather}i \pd_t\left(
\begin{array}{c}
 \psi  \\
 \chi  \\
\end{array}
\right)=\left(
\begin{array}{cc}
 \tilde{H} & \mu  \hat{G}^{n-1} \\
 \mu \hat{G}^{n-1} & -\mu^2 \hat{G}^{n-2} \Sigma \\
\end{array}
\right) \left(
\begin{array}{c}
 \psi  \\
 \chi  \\
\end{array}
\right).
\end{gather}

This shows that, by extending the state function, one can reduce the order of the spatial derivative entering the Hamiltonian from $n$ to $(n - 1)$.
It is then seen, by induction, that one can also reduce it to the first order. In the limit $\mu \to \infty$, the additional dispersion branches
that emerge due to this procedure correspond to infinite frequencies and thus cannot affect the dynamics of $\psi $. 

\subsection{Example}

Let us consider a simple case with $n = 2$ as an example. Specifically, suppose
\begin{gather}
\hat{\msf{H}} = g^2 \hat{k}^2 + \hat{\varrho},
\end{gather}
where $g$ is a real constant, and $\hat{\varrho} = o(\hat{k}^2)$. Then, $\hat{K}$ is a \textit{scalar} operator, $\hat{K} = \hat{G} = g \hat{k}$. This leads to the following equation in the extended space,
\begin{gather}i \pd_t\left(
\begin{array}{c}
 \psi  \\
 \chi  \\
\end{array}
\right)=\left(
\begin{array}{cc}
 \hat{\varrho} & \mu  g  \hat{k} \\
 \mu  g  \hat{k} & -\mu ^2 \\
\end{array}
\right) \left(
\begin{array}{c}
 \psi  \\
 \chi  \\
\end{array}
\right).\label{eq:extspb}
\end{gather}

When $\hat{\varrho}$ is homogeneous, \Eq{eq:extspb} has eigenvectors
\begin{gather}\left(
\begin{array}{c}
 \psi  \\
 \chi  \\
\end{array}
\right)=e^{-i \omega t+ i k x}\left(
\begin{array}{c}
 \psi_0 \\
 \chi_0 \\
\end{array}
\right).
\end{gather}
where $\psi_0$ and $\chi_0$ are complex constants. The dispersion relation connecting $\omega$ with $k$ is
\begin{gather}\det \left(
\begin{array}{cc}
 \varrho(k) - \omega  & \mu g k \\
 \mu g k & - \mu^2 - \omega  \\
\end{array}
\right) = 0,
\end{gather}
where $\varrho(k)$ is the eigenvalue of $\hat{\varrho}$. This leads to
\begin{gather}\notag
\omega_{1,2} = \frac{\varrho(k) - \mu^2}{2} \pm \frac{1}{2} \sqrt{[\mu^2-\varrho (k)]^2 + 4\mu^2\msf{H}(k)},
\end{gather}
where $\msf{H}(k)$ is the corresponding eigenvalue of $\hat{\msf{H}}$. In the limit $\mu \to \infty$, one of these solutions approaches $\msf{H}(k)$, as intended, whereas the other one corresponds to infinite frequency and thus can be ignored.

\section{Auxiliary calculations}

Here we present auxiliary calculations that are used in \Sec{sec:dirac} to derive $U$ for a Dirac particle. 

\subsection{Expressions for $\boldsymbol{\mc{P}_t}$, $\boldsymbol{\mc{P}_x}$, $\boldsymbol{\mc{Q}_t}$, and $\boldsymbol{\mc{Q}_x}$}
\label{app:auxdirac}

Using that
\begin{gather}\notag
\frac{\pd \Xi}{\pd p_j} = \sqrt{\frac{m + \varepsilon}{2\varepsilon}} \, \frac{\pd}{\pd p_j} \left(
\begin{array}{c}
 \mathbb{I}_2 \\
 \frac{\vec{\sigma} \cdot \vec{p}}{m + \varepsilon} \\
\end{array}
\right) 
= \sqrt{\frac{m + \varepsilon}{2\varepsilon}}\left(
\begin{array}{c}
 0 \\
 \frac{\sigma^j}{m + \varepsilon} \\
\end{array}
\right),
\end{gather}
we obtain the following expressions for $\mc{P}_t$ and $\mc{P}_x$:
\begin{align}
\mc{P}_t 	& = \text{Im}\left(\Xi^\dag \frac{\pd \Xi}{\pd p_j} \, \pd_t p_j\right) \notag \\
				& = \text{Im} \left[ \frac{m + \varepsilon}{2\varepsilon} \left(
							\begin{array}{cc} \mathbb{I}_2 & \frac{\vec{\sigma} \cdot \vec{p}}{m + \varepsilon} \\
							\end{array}
						\right) \left(
							\begin{array}{c}
 							0 \\ \frac{\sigma^j}{m + \varepsilon} \\
							\end{array}
						\right)\pd_tp_j \right]\notag \\
				& = \text{Im} \left[
				\frac{(\vec{\sigma} \cdot \vec{p})(\vec{\sigma} \cdot \pd_t \vec{p})}{2\varepsilon (m + \varepsilon)}
				\right]
								 \notag \\
				& = \text{Im} \left[\frac{(\pd_t \vec{p}) \cdot \vec{p} + i(\vec{p} \times \pd_t \vec{p})\cdot \vec{\sigma}}{2\varepsilon (m + \varepsilon)} 
    						\right] \notag \\
				& = \frac{\vec{v} \times \pd_t \vec{p}}{2(m + \varepsilon)} \cdot \vec{\sigma},
\end{align}
\begin{align}
\mc{P}_x	& = \text{Im} \left(\Xi^\dag \alpha^i \, \frac{\pd \Xi}{\pd p_j}\,\pd_i p_j\right) \notag \\
				& = \text{Im} \left[ \frac{m + \varepsilon}{2\varepsilon}\left(
							\begin{array}{cc}
 								\mathbb{I}_2 & \frac{\vec{\sigma} \cdot \vec{p}}{m + \varepsilon} \\
							\end{array}
						\right) \left(
							\begin{array}{cc}
 								0 & \sigma^i \\
 								\sigma^i & 0 \\
							\end{array}
						\right) \left(
							\begin{array}{c}
 								0 \\
 								\frac{\sigma^j}{m + \varepsilon} \\
								\end{array}
						\right) \pd_i p_j \right]\notag \\
				& = \text{Im} \left[ \frac{m + \varepsilon}{2\varepsilon} \left(
						\begin{array}{cc}
 							\mathbb{I}_2 & \frac{\vec{\sigma} \cdot \vec{p}}{m + \varepsilon} \\
						\end{array}
						\right) \left(
							\begin{array}{c}
 								\frac{\sigma^i \sigma^j}{m + \varepsilon} \\
 							0 \\
							\end{array}
						\right) \pd_i p_j \right]\notag \\
				& = \text{Im} \left[ \frac{(\vec{\sigma} \cdot \del)(\vec{\sigma} \cdot \vec{p})}{2\varepsilon} \right] \notag \\
				& = \text{Im} \left[ \frac{(\del \cdot \vec{p}) + i(\del \times \vec{p}) \cdot \vec{\sigma}}{2\varepsilon} \right] \notag \\
				& = \frac{1}{2\varepsilon} \, (\del \times \vec{p}) \cdot \vec{\sigma} \notag \\
				& = -\frac{q}{2\varepsilon} \, (\del \times \vec{A}) \cdot \vec{\sigma} \notag \\
				& = -\frac{q \vec{B}}{2\varepsilon} \cdot \vec{\sigma}.
\end{align}

Using that
\begin{align}
\frac{\pd \Xi}{\pd \varepsilon} & = \left(
									\begin{array}{c}
		 									\mathbb{I}_2 \\
 											\frac{\vec{\sigma} \cdot \vec{p}}{m + \varepsilon} \\
									\end{array}
									\right)
									\frac{\pd}{\pd \varepsilon} \, \sqrt{\frac{m + \varepsilon}{2\varepsilon}}
									+\sqrt{\frac{m + \varepsilon}{2\varepsilon}} \, \frac{\pd}{\pd \varepsilon}\left(
									\begin{array}{c}
			 							\mathbb{I}_2 \\
 										\frac{\vec{\sigma} \cdot \vec{p}}{m + \varepsilon} \\
									\end{array}
									\right) \notag \\
						& = \sqrt{\frac{m + \varepsilon}{2\varepsilon}} \left[
								-\frac{m}{2\varepsilon (m + \varepsilon)}
								\left(
								\begin{array}{c}
 									\mathbb{I}_2 \\
 									\frac{\vec{\sigma} \cdot \vec{p}}{m + \varepsilon} \\
								\end{array}
								\right)
								- \left(
								\begin{array}{c}
 									0 \\
 									\frac{\vec{\sigma} \cdot \vec{p}}{(m + \varepsilon)^2} \\
								\end{array}
								\right)
								\right] \notag \\
						& = -\frac{m \Xi}{2\varepsilon (m + \varepsilon)}
								-\sqrt{\frac{m + \varepsilon}{2\varepsilon}}
								\begin{pmatrix}
 								0 \\
 								\frac{\vec{\sigma} \cdot \vec{p}}{(m + \varepsilon)^2} \\
								\end{pmatrix},\notag
\end{align}
we also obtain the following expressions for $\mc{Q}_t$ and $\mc{Q}_x$:
\begin{align}
\mc{Q}_t = &\,	\text{Im} \left(\Xi^\dag \frac{\pd \Xi}{\pd \varepsilon} \, \pd_t \varepsilon \right) \notag \\
				 = &	-\text{Im} \left[ \frac{m \pd_t \varepsilon}{2\varepsilon (m + \varepsilon)}\, \Xi^\dag \Xi \right] \notag \\
				   &  -\text{Im} \left[ \frac{m + \varepsilon}{2\varepsilon} 
							\begin{pmatrix}
								 \mathbb{I}_2 & \frac{\vec{\sigma} \cdot \vec{p}}{m + \varepsilon} 
							\end{pmatrix}									
							\begin{pmatrix}
								0 \\
 								\frac{\vec{\sigma} \cdot \vec{p}}{(m + \varepsilon)^2} \\
							\end{pmatrix}
						\right] \notag\\
				= & -\text{Im} \left[ \frac{m \pd_t \varepsilon}{2\varepsilon (m + \varepsilon)}\, \mathbb{I}_2 \right]
					  -\text{Im} \left[ 
							\frac{(\vec{\sigma} \cdot \vec{p})(\vec{\sigma} \cdot \vec{p})  }{2\varepsilon (m+\varepsilon)^2 }\, 
							\pd_t \varepsilon  \right] \notag \\
				= & -\text{Im} \left[ 
							\frac{\vec{p}^2}{2\varepsilon (m+\varepsilon)^2 } \,
							\pd_t \varepsilon  \right] \notag \\
				= &\, 0,
\end{align}
\begin{align}
\mc{Q}_x = &\, \text{Im} \left(\Xi^\dag \alpha^i\frac{\pd \Xi}{\pd \varepsilon} \, \pd_i\varepsilon \right) \notag \\
				= & -\text{Im} \left[\frac{m}{2\varepsilon (m + \varepsilon)} \, \Xi^\dag \alpha^i \Xi \pd_i \varepsilon \right] \notag \\
				  & -\text{Im} \left[ \frac{m + \varepsilon}{2\varepsilon} 
									\begin{pmatrix}
								 		\mathbb{I}_2 & \frac{\vec{\sigma} \cdot \vec{p}}{m + \varepsilon} 
									\end{pmatrix}	
									\begin{pmatrix}
										0 & \sigma^i \\	
										\sigma^i & 0 			
									\end{pmatrix}																	
									\begin{pmatrix}
										0 \\
 										\frac{\vec{\sigma} \cdot \vec{p}}{(m + \varepsilon)^2} \\
									\end{pmatrix}
									\pd_i \varepsilon 
									\right] \notag\\				
				= & -\text{Im} \left[ \frac{m}{2\varepsilon (m + \varepsilon)} \, \mathbb{I}_2 (\vec{v}_0 \cdot \del) \varepsilon \right] \notag \\
				  &	-\text{Im} \left[
										 \frac{ (\vec{\sigma} \cdot \del \varepsilon) (\vec{\sigma} \cdot \vec{p}) }{2\varepsilon(m + \varepsilon)} \right]  \notag\\
				= &\, \frac{\vec{v} \times \del \varepsilon}{2(m + \varepsilon)} \cdot \vec{\sigma}.
\end{align}

\subsection{Anomalous magnetic moment}
\label{app:diracan}

Here, we show that
\begin{widetext}
\begin{align}
\psi^\dag \beta \sigma_{\mu\nu} F^{\mu\nu} \psi = &\, \frac{m + \varepsilon}{\varepsilon}\, a^\dag
																			\begin{pmatrix}
																				\mathbb{I}_2 & 
																				\frac{\vec{\sigma} \cdot \vec{p}}{m + \varepsilon}
																			\end{pmatrix}
																			\begin{pmatrix}
																				\mathbb{I}_2 & 0 \\
																				0 & -\mathbb{I}_2
																			\end{pmatrix}
																			\begin{pmatrix}
																				- \vec{\sigma} \cdot \vec{B} & i  \vec{\sigma} \cdot \vec{E}  \\
																				i \vec{\sigma} \cdot \vec{E} & -  \vec{\sigma} \cdot \vec{B}  \\
																			\end{pmatrix}
																			\begin{pmatrix}
																				\mathbb{I}_2 \\
																				\frac{\vec{\sigma} \cdot \vec{p}}{m + \varepsilon}
																			\end{pmatrix}
																			a \notag \\
							= & - \frac{m + \varepsilon}{\varepsilon} \, a^\dag
									\begin{pmatrix}
										\mathbb{I}_2 & 
										\frac{\vec{\sigma} \cdot \vec{p}}{m + \varepsilon}
									\end{pmatrix}
									\begin{pmatrix}
										\vec{\sigma} \cdot \vec{B} - i \vec{\sigma} \cdot \vec{E}\, \frac{\vec{\sigma} \cdot \vec{p}}{m + \varepsilon} \\
										i \vec{\sigma} \cdot \vec{E} -  \vec{\sigma} \cdot \vec{B}\, \frac{\vec{\sigma} \cdot \vec{p}}{m + \varepsilon}
									\end{pmatrix}
									a \notag \\
							= & - \frac{m + \varepsilon}{\varepsilon} \, a^\dag
									\begin{pmatrix}
										\mathbb{I}_2 & 
										\frac{\vec{\sigma} \cdot \vec{p}}{m + \varepsilon}
									\end{pmatrix}
									\begin{pmatrix}
										\vec{B} \cdot \vec{\sigma} - \frac{i\vec{E} \cdot \vec{p}}{m + \varepsilon} 
												+ \frac{(\vec{E} \times \vec{p}) \cdot \vec{\sigma}}{m + \varepsilon}  \\
										i \vec{\sigma} \cdot \vec{E} - \frac{\vec{B} \cdot \vec{p}}{m + \varepsilon}
												- \frac{i(\vec{B} \times \vec{p}) \cdot \vec{\sigma}}{m + \varepsilon}
									\end{pmatrix}
									a \notag \\
							= & 	- \frac{m + \varepsilon}{\varepsilon} \, a^\dag
									\left[
							 			\vec{B} \cdot \vec{\sigma} - \frac{i \vec{E} \cdot \vec{p}}{m + \varepsilon}
							 			+ \frac{(\vec{E} \times \vec{p}) \cdot \vec{\sigma}}{m + \varepsilon}  
							 			+ \frac{i \vec{\sigma} \cdot \vec{p}}{m + \varepsilon} \vec{\sigma} \cdot \vec{E} 
							 			- \frac{\vec{\sigma} \cdot \vec{p}}{m + \varepsilon} \, \frac{\vec{B} \cdot \vec{p}}{m + \varepsilon}
							 			- \frac{i \vec{\sigma} \cdot\vec{p}}{m+\varepsilon}\,\frac{(\vec{B}\times \vec{p}) \cdot \vec{\sigma}}{m+ \varepsilon}
									\right]
									a\notag \\
							= & 	- \frac{m + \varepsilon}{\varepsilon} \, a^\dag
									\left[
							 			\vec{B} \cdot \vec{\sigma} + \frac{(\vec{E}\times \vec{p}) \cdot \vec{\sigma}}{m + \varepsilon}  
							 			- \frac{(\vec{p}\times \vec{E}) \cdot \vec{\sigma}}{m + \varepsilon} 
							 			- \frac{\vec{\sigma} \cdot \vec{p}}{m + \varepsilon} \, \frac{\vec{B} \cdot \vec{p}}{m + \varepsilon}
							 			- \frac{i \vec{\sigma} \cdot \vec{p}}{m + \varepsilon} \, \frac{(\vec{B}\times \vec{p}) \cdot \vec{\sigma}}{m + \varepsilon}
									\right] 
									a \notag \\
							= & 	- \frac{m + \varepsilon}{\varepsilon} \, a^\dag
									\left[
							 			\vec{B} \cdot \vec{\sigma} - \frac{2(\vec{p}\times \vec{E}) \cdot \vec{\sigma}}{m + \varepsilon}
							 			- \frac{\vec{\sigma} \cdot \vec{p}}{m + \varepsilon} \, \frac{\vec{B} \cdot \vec{p}}{m + \varepsilon}
							 			+ \frac{1}{m + \varepsilon} \, \frac{(\vec{p} \times (\vec{B}\times \vec{p})) \cdot \vec{\sigma}}{m + \varepsilon}
									\right]
									a \notag \\
							= & 	- \frac{m + \varepsilon}{\varepsilon} \, a^\dag
									\left[
							 			\vec{B} \cdot \vec{\sigma} - \frac{2(\vec{p}\times \vec{E}) \cdot \vec{\sigma}}{m + \varepsilon}  
							 			- \frac{\vec{\sigma} \cdot \vec{p}}{m + \varepsilon} \, \frac{\vec{B} \cdot \vec{p}}{m + \varepsilon}
							 			+ \frac{1}{m + \varepsilon} \, \frac{\vec{p}^2 (\vec{B} \cdot \vec{\sigma}) 
							 			- (\vec{p} \cdot\vec{B)(\vec{p} \cdot\vec{\sigma})}}{m + \varepsilon}
										\right]
									a\notag \\
							= & - \frac{m + \varepsilon}{\varepsilon} \, a^\dag
									\left[
							 			\vec{B} \cdot \vec{\sigma} -  2\, \frac{(\vec{p}\times \vec{E}) \cdot \vec{\sigma}}{m + \varepsilon}  
							  			- 2\, \frac{\vec{\sigma} \cdot \vec{p}}{m + \varepsilon} \, \frac{\vec{B} \cdot \vec{p}}{m + \varepsilon}
							  			+ \frac{\vec{p}^2 (\vec{B} \cdot \vec{\sigma})}{(m + \varepsilon)^2}
									\right]
									a \notag \\
							= & - a^\dag 
									\left[ \frac{\gamma + 1}{\gamma} \, \vec{B} \cdot \vec{\sigma}
							  			- 2(\vec{v} \times \vec{E}) \cdot \vec{\sigma}  
							  			- \frac{2\gamma}{\gamma + 1} \, (\vec{\sigma} \cdot \vec{v}) (\vec{B} \cdot \vec{v})
							  			+ \frac{\gamma}{\gamma + 1}\, \vec{v}^2 (\vec{B} \cdot \vec{\sigma}) 
									\right]
									a \notag \\
							= & - 2 a^\dag
										\left[
											\vec{\sigma} \cdot \vec{B} 
											- (\vec{v} \times \vec{E}) \cdot \vec{\sigma} 
											- \frac{\gamma}{\gamma + 1} \, (\vec{\sigma} \cdot \vec{v}) (\vec{B} \cdot \vec{v})
										\right] a.
\end{align}
\end{widetext}



\begin{thebibliography}{10}
\bibitem{foot:history}
In addition to the already mentioned papers, this issue is discussed in
  \Sec{sec:discussion} and also in \Sec{sec:comparison}.
\bibitem{foot:nege}
The sign depends on the metric signature in $\psi$ space, which, in some cases,
  can be not positive-definite. For further details, see \Ref{my:wkin}.
\bibitem{foot:gauge}
We call $-U$ a gauge Hamiltonian because it emerges as a result of the variable
  transformation ${\psi \mapsto a}$, which is a (projected) gauge
  transformation.
\bibitem{foot:real}
More rigorously, the independent variables should be $\text{Re}\,z$ and
  $\text{Im}\,z$ \cite{my:wkin}. Then $\mu$ and $\nu$ are real by definition.
\bibitem{foot:debat}
The question whether the procedure in \Ref{ref:barut85} is merely a
  mathematical exercise or actually describes particles as they are understood
  commonly is something that was not quite answered in the paper.
\bibitem{foot:kinfl}
One can draw a parallel here with the hydrodynamic description of collisionless
  plasma. As formal corollaries of kinetic theory, hydrodynamic equations are
  exact. But they are meaningless without a proper closure, and that is
  derivable only in limiting cases. Thus, keeping a complete (infinite in this
  case) set of hydrodynamic equations is never warranted; either a reduced
  model is possible, or no hydrodynamic description is possible at all.
\bibitem{foot:prob}
This is particularly important for applying our general theory to classical
  waves (\Sec{sec:discussion}), for which the probability interpretation is,
  basically, irrelevant.
\end{thebibliography}
\end{document}